\documentclass[reprint,aps,prd,twocolumn,notitlepage,nofootinbib,showpacs,preprintnumbers,superscriptaddress]{revtex4-1}
\usepackage[T1]{fontenc}
\usepackage[utf8]{inputenc}
\usepackage{bm}
\usepackage{amsmath}
\usepackage{amssymb}
\usepackage{mathrsfs}
\usepackage{esint}
\usepackage{xcolor}
\usepackage{enumitem}
\usepackage{multirow}
\usepackage{graphicx}
\PassOptionsToPackage{normalem}{ulem}
\usepackage{ulem}
\usepackage{times}
\usepackage{hyperref}

\begin{document}
\preprint{YITP-23-09, IPMU23-0003}

\title{Minimally modified gravity with auxiliary constraints formalism}

\author{Zhi-Bang Yao}%
      \email[Email: ]{yaozhb@mail2.sysu.edu.cn}
      \affiliation{School of Physics and Astronomy, Sun Yat-sen University, Zhuhai 519082, China}

\author{Michele Oliosi}%
\email[Email: ]{michele.oliosi@yukawa.kyoto-u.ac.jp}   
\affiliation{Center for Gravitational Physics and Quantum Information, Yukawa Institute for Theoretical
Physics,~\\
 Kyoto University, Kyoto 606-8502, Japan}     

\author{Xian Gao}%
    \email[Email: ]{gaoxian@mail.sysu.edu.cn}
    \affiliation{School of Physics and Astronomy, Sun Yat-sen University, Zhuhai 519082, China}

\author{Shinji Mukohyama}%
    \email[Email: ]{shinji.mukohyama@yukawa.kyoto-u.ac.jp
    }
\affiliation{Center for Gravitational Physics and Quantum Information, Yukawa Institute for Theoretical
Physics,~\\
 Kyoto University, Kyoto 606-8502, Japan}
\affiliation{Kavli Institute for the Physics and Mathematics of the Universe (WPI),
~\\
The University of Tokyo Institutes for Advanced Study,~\\
The University of Tokyo, Kashiwa, Chiba 277-8583, Japan}

\date{May 30, 2023}

\begin{abstract}
We investigate the possibility of reducing the number of degrees of freedom (d.o.f.) starting from generic metric theories of gravity by introducing multiple auxiliary constraints (ACs), under the restriction of retaining spatial covariance as a gauge symmetry. Arbitrary numbers of scalar-, vector- and tensor-type ACs are considered a priori, yet we find that no vector- and tensor-type constraints should be introduced, and that scalar-type ACs should be no more than four for the purpose of constructing minimally modified gravity (MMG) theories which propagate only two tensorial d.o.f., like general relativity (GR). Through a detailed Hamiltonian analysis, we exhaust all the possible classifications of ACs and find out the corresponding minimalizing and symmetrizing conditions for obtaining the MMG theories. In particular, no condition is required in the case of four ACs, hence in this case the theory can couple with matter consistently and naturally. To illustrate our formalism, we build a concrete model for this specific case by using the Cayley-Hamilton theorem and derive the dispersion relation of the gravitational waves, which is subject to constraints from the observations.
\end{abstract}

\maketitle

\section{Introduction \label{Sec_Intro}}

After the first binary black hole coalescence event, GW150914, detected
by LIGO in 2015 \citep{Abbott2016}, there have been more than fifty
compact binary merger events reported by the LIGO-Virgo Scientific
Collaborations (LVSC) \citep{Abbott2019,Abbott2021} heralding the
era of gravitational wave (GW) for astrophysics and cosmology which
provides the first-ever window to explore the nature of gravity in
the strong-field regime. So far, GR stands strongly against tests
such as consistency checks \citep{Abbott2021a}, merger remnants \citep{Healy2014}
and the properties of the generation and propagation (e.g. the propagating
speed, dispersion relations and polarization states) of GWs \citep{Abbott2016a,Yunes:2016jcc,Cai2017,Samajdar2017,Abbott2017c,Abbott2017,Abbott2017a,Abbott2017b}
(see \citep{Krishnendu2021} for reviews).

In particular, due to the spacetime diffeomorphism symmetry, GR predicts
that only two transverse-traceless tensor modes of GWs are propagating.
This is a significant feature to distinguish GR from usual modified
gravity theories \citep{Clifton:2011jh,Will2014}, in which additionally
non-tensorial polarization mode(s) \citep{Arzoumanian2020,Chen2021,Arzoumanian2021}
could be propagating as well. For instance, there is a scalar mode
beyond the two tensor modes in the $f\left(R\right)$ and generic
scalar-tensor (ST) theories \citep{Brans:1961sx,Horndeski:1974wa,ArmendarizPicon1999,Chiba:1999ka,Deffayet:2011gz,Langlois:2015cwa}
(see also \citep{Liang2017,Gong2018a,Hou2018}), the vector modes
appear in the vector-tensor theories (e.g. the Einstein-Æther theory
\citep{Jacobson2004}) while both scalar and tensor modes could be
excited simultaneously in addition to the tensor modes in scalar-vector-tensor
theories, such as the TeVeS theory \citep{Bekenstein2004}. Therefore,
people have put a lot of effort into extracting information on the
polarization states from GWs signals \citep{Gong2018,Hagihara2018,Takeda2019,Takeda2021,Zhang2022}
for the purpose of falsifying gravity theories via polarization states
of GWs. However, limited by the orientation of the detectors arranged
at LVSC, so far the best we have known about the polarization states
from the GWs signals is that the purely tensor polarization is strongly
favored over the purely scalar or vector polarizations \citep{Abbott2017c}.
In light of the establishment of the network of ground-based detectors
including advanced LIGO, Virgo \citep{LIGOScientific:2014pky,TheVirgo:2014hva,Abbott2018},
KAGRA \citep{Somiya2012,Aso2013} and LIGO India, more information
will be available and we may then have the probing ability for separating
polarization modes in the near future.

Naturally, in this context there comes a crucial question, that is,
whether GR is the unique theory with two tensorial degrees of freedom
(TTDOF) of gravity. According to the Lovelock's theorem \citep{Lovelock:1970,Lovelock1972},
GR is the unique theory in which the metric is the only field in the
gravity sector and obeys second order equations of motion with general
covariance and locality in the four dimensional spacetime and therefore
GR is indeed the unique TTDOF gravity theory when the assumptions
of Lovelock's theorem are preserved.

However, this uniqueness does not hold anymore when these assumptions
are relaxed to some extent \footnote{Recently, the authors in \citep{Glavan2020} attempted to circumvent
the Lovelock's theorem to construct an alternative gravity theory
with TTDOF by rescaling the coupling constant of the Gauss-Bonnet
term and therefore non-trivially modifying the Einstein's field equation
in four dimensional spacetime. However, the Lovelock's theorem is
a statement about the equations of motion (but not just about the
action) and therefore directly excludes such a possibility. Indeed,
it was soon realized that taking the limit of Gauss-Bonnet term from
higher dimension to 4 dimension is path-dependent which should be
regularized properly and, as a consequence, additional d.o.f. appears
in the regularized 4D Einstein-Gauss-Bonnet theories (see \citep{Fernandes2022}
for reviews). Alternatively, by breaking the 4D diffeomorphism invariance
down to the 3D spatial diffeomorphism invariance, one can construct
a consistent 4D Einstein-Gauss-Bonnet theory with TTDOF \citep{Aoki:2020lig,Aoki:2020iwm,Aoki:2020ila}.
We will briefly introduce this theory in Appendix~\ref{exam_EGB}.}. In fact, an alternative gravity theory with TTDOF was proposed and
dubbed the Cuscuton in 2007 \citep{Afshordi2007} by introducing an
incompressible (i.e. propagating at infinite sound speed) scalar field
$\phi\left(t,\vec{x}\right)$ with a non-vanishing and timelike vacuum
expectation value of its first derivative. The scalar d.o.f. with
infinite sound speed in the Cuscuton manifests itself as an instantaneous
mode. As discussed in \citep{DeFelice:2018ewo,DeFelice:2021hps} and
\citep{DeFelice:2022uxv} for a generalized instantaneous mode called
a shadowy mode in the context of U-DHOST and VCDM respectively, such
a mode with infinite sound speed satisfies an elliptic equation on
each constant-$\phi$ hypersurface. In the so-called unitary gauge,
i.e. with $\phi=\phi\left(t\right)$ \citep{ArkaniHamed:2003uy},
these hypersurfaces on which the elliptic equation is defined agree
with time slices and thus the equation of motion for the instantaneous/shadowy
mode does not include time derivatives, meaning that the Cuscuton
only propagates the TTDOF \citep{Gomes2017}. In other choices of
time slicing, the equation of motion for the instantaneous/shadowy
mode includes time derivatives but is still elliptic. Therefore by
imposing a proper boundary condition the instantaneous/shadowy mode
is uniquely determined by other degrees of freedom. As a result the
Cuscuton propagates TTDOF only, irrespective of whether the unitary
gauge is adopted or not (see \citep{DeFelice:2018ewo,DeFelice:2021hps}
and \citep{DeFelice:2022uxv} for corresponding discussions in U-DHOST
and VCDM ).

Generally, in the unitary gauge, general covariance is broken into
the spatial covariance therefore the Cuscuton (or a generic ST theory)
can be classified as a spatially covariant (SC) framework \citep{Gao:2014soa}
and inversely the ST theory can be also recovered from a SC theory
by introducing a Stueckelberg field \citep{Gao2020b,Gao2020a,Hu2022}.
Due to the reduction of symmetry, in addition to the two tensorial
d.o.f., the SC gravity theory with a nondynamical lapse function propagates
one scalar d.o.f. \citep{Gao:2014fra}. Nevertheless, a class of TTDOF
gravity theories was proposed within a special SC framework where
the lapse function enters the Lagrangian linearly. The resulting theory
was dubbed the MMG theory \citep{Lin:2017oow}, which indicates that
GR is modified without changing its d.o.f.. Similarly, a class of
TTDOF theories was identified within the more general ST theory with
higher order derivatives under the unitary gauge \citep{Iyonaga2018}.
In the case of a nondynamical lapse function, the general conditions
to eliminate the scalar d.o.f. were found in \citep{Gao:2019twq,Hu:2021yaq}.
This idea was also applied to the more general SC framework with a
dynamical lapse function that enters the Lagrangian nonlinearly \citep{Gao2019}.
The conditions to have TTDOF in the presence of a dynamical lapse
function were analyzed in \citep{Lin2021}.

According to the above works, within the general SC framework, MMG
theories exist as long as some additional conditions are satisfied,
which we dub the TTDOF conditions. The example of the first TTDOF
condition arose in \citep{Iyonaga2018}, which we dubbed the degeneracy
condition to indicate that the sector of the lapse function and extrinsic
curvature is degenerate. The self-consistency condition identified
in \citep{Lin:2017oow}, which we dub the second TTDOF condition,
is used to prevent the number of phase space dimensions from being
odd.

The TTDOF conditions are nonlinear functional differential equations
of the Lagrangian, which are difficult to be solved in general. We
are only able to solve them with some particular ansatz of actions,
such as, the square root gravity in \citep{Lin:2017oow}, the extended
Cuscuton in \citep{Iyonaga2018} and the quadratic extrinsic curvature
(QEC) gravity in \citep{Gao:2019twq} (see the cosmological constraints
to this model in \citep{Iyonaga2021,Hiramatsu2022}). Another difficulty
of MMG theories is the problem of coupling with matter consistently,
which happens when there are extra first-class constraint(s) (other
than the original six first-class constraints associated with the
spatial diffeomorphism) appearing in the MMG theories \citep{Carballo-Rubio2018}.
More precisely, the extra first-class constraint(s) would be downgraded
to be second-class when the theory is naively coupled with Lorentz
covariant matter, at which point the suppressed scalar d.o.f. arises
again.

The resolution of these difficulties arising for MMG theories become
more transparent in the phase space than in the configuration space.
This is because counting the number of d.o.f. is transparent by means
of an explicit constraint analysis in the phase space. For this reason,
a number of works have utilized the Hamiltonian approach. For instance,
by imposing the linearity of the lapse function in the Hamiltonian
\citep{Mukohyama2019} instead of the Lagrangian \citep{Lin:2017oow},
the self-consistency condition was reformulated in a much simpler
expression by solving the simplified condition. In these works, the
$f\left(\mathcal{H}\right)$ theory which is a particular MMG theory
constructed in \citep{Carballo-Rubio2018} was rediscovered and the
``kink'' model based on the $f\left(\mathcal{H}\right)$ theory
was shown later to fit the Planck data better than the $\Lambda$CDM
model \citep{Aoki2020b}. The matter-coupling problem can also be
addressed in the phase space. As another instance, in \citep{Aoki2018,Aoki:2018brq,DeFelice2020b},
the authors introduce the so-called gauge fixing condition to the
Hamiltonian of the MMG theory obtained by performing a canonical transformation
of GR. As a result, the first-class constraint associated with temporal
diffeomorphism is split into a pair of second-class constraints, which
allow the theory to couple with the Lorentz covariant matter consistently.
Another proposal for addressing the same problem is adopted in \citep{Carballo-Rubio2018,Lin2019,Lin2019a},
where the additional first-class constraint is maintained by modifying
the Hamiltonian constraint of the matter sector such that the constraints
algebra is kept closed. However, as a price, matter no longer behaves
in the usual Lorentz covariant manner.

Inspired by the above works, a more straightforward approach to constructing
the MMG theories was proposed in \citep{Yao2021}, where the so-called
(scalar-type) auxiliary constraint (AC) is introduced to a general
total Hamiltonian respecting the spatial diffeomorphism with a non-dynamical
lapse function. The Hamiltonian carries two tensorial and one scalar
d.o.f. at the beginning. The AC is used to constrain the trajectories
of canonical variables such that the unwanted scalar d.o.f. is suppressed.
It is therefore introduced to assist in locating the MMG theories
in the space of theories and is eventually part of the definition
of the theory. This AC can be also considered as the generalization
of the gauge-fixing condition that addresses the matter-coupling problem
mentioned above. Nevertheless, the phase space constrained by the
AC is still insufficient to ensure a MMG theory because the AC is
introduced via a generic function of the canonical variables. Generally,
additional TTDOF conditions are still needed, which are renamed as
the ``minimalizing conditions'' in \citep{Yao2021}, to underline
that they are the conditions that ``minimalize'' the space of more
general theories into the MMG theory space. Even though the AC is
initially introduced by hand, we emphasize that it is actually nothing
but one possible kind of constraint structure for the MMG theories
and just part of the definition of the theory. In principle there
could be more than one AC, which can be thought of as yet unrevealed
territory among the constraint structures of the MMG theories.

In this work, we are going to complete the constraint analysis for
the MMG theories with multiple ACs. In principle, not only the scalar-type
but also the vector- and the tensor-type ACs might possibly appear.
Hence we will firstly investigate the possibilities of minimizing
the number of d.o.f. by introducing an arbitrary number of any of
the above types of ACs to a general total Hamiltonian, which still
preserves spatial diffeomorphism invariance. Throughout, it is also
important to limit the number of ACs, otherwise the system will be
over-constrained thus become physically inconsistent. In order to
determine the maximum number of each type of ACs, we will first assume
that they are all classified as second-class then count the number
of d.o.f. with respect to an arbitrary background via the Hamiltonian
analysis. By requiring that the number of each type of d.o.f.~be
non-negative, we will find limits on the number of each type of ACs.
In fact, as we will see in the next section, the number of scalar-type
ACs should not exceed four and no vector- and tensor-type ACs are
allowed, because the phase space in the vector sector is already sufficiently
constrained by the spatial diffeomorphism constraints, while the tensor
sector should not be constrained in order to have the correct number
of d.o.f.

Following this, we will construct a consistent SC framework with multiple
(scalar-type) ACs as our starting point for searching the MMG theories.
According to the number of introduced ACs, we will divide the theories
into four cases and for the purpose of obtaining the MMG theories,
the scalar d.o.f. should be completely eliminated by the primary and
secondary constraints, which will divide the ACs into different classifications
for each case. To further contrive the classification, the ACs and
the canonical Hamiltonian should satisfy not only the corresponding
minimalizing conditions mentioned above but also symmetrizing conditions
which are the sufficient but not necessary conditions to end up with
a MMG theory and are required to enhance the gauge symmetries of the
theory. As a results, we will exhaust all the possible constraint
structures for the MMG theories with multiple ACs, thus leading to
a complete classification.

To illustrate this formalism, we will construct a concrete model with
four ACs by using the Cayley-Hamilton theorem. This theory can couple
with matter consistently without further conditions. We will investigate
the dispersion relation of tensor perturbations around a flat FLRW
background, and show how some coefficients of the theory can be constrained
by the observations.

The rest of the current paper is organized as follows. In sec. \ref{Sec_ACF_MAC},
we investigate the possibilities of reducing the number of d.o.f.
by introducing different types of ACs and determine a consistent general
framework with multiple ACs as our starting point for searching the
MMG theories. In sec. \ref{Sec_MMG_MAC}, we find the minimalizing
and symmetrizing conditions for each class of MMG theories, each of
which is described in subsections \ref{Subsec_S4}-\ref{Subsec_S1}.
As an illustrative example, we construct a concrete model with four
ACs and study the modified dispersion relation of the GWs by performing
a tensor perturbation in sec. \ref{Sec_CH_FAC}. Finally, we conclude
this work in sec. \ref{Sec_cld}.

\section{A consistent framework with auxiliary constraints \label{Sec_ACF_MAC}}

In this section, we will investigate the possibilities of reducing
the number of d.o.f. by introducing multiple primary ACs and construct
a consistent framework as our starting point for searching for the
MMG theories. We will adopt the Arnowitt-Deser-Misner (ADM) formalism
in which the lapse function, shift vector, induced metric and their
conjugate momenta are denoted by $\left\{ N,N^{i},h_{ij};\pi,\pi_{i},\pi^{ij}\right\} $
respectively and $\nabla_{i}$ is the spatially covariant derivative
compatible with $h_{ij}$. Without loss of generality, we start with
the following general total Hamiltonian 
\begin{eqnarray}
H_{\text{T}} & = & \int\text{d}^{3}x\Big[\mathscr{H}\left(N,\pi,h_{ij},\pi^{ij};\nabla_{i}\right)+N^{i}\mathcal{H}_{i}\nonumber \\
 &  & +\lambda^{i}\pi_{i}+\mu_{\text{n}}\mathcal{S}^{\text{n}}+\nu_{\text{m}}^{i}\mathcal{V}_{i}^{\text{m}}+\rho_{\text{r}}^{ij}\mathcal{T}_{ij}^{\text{r}}\Big],\label{Int_HT_anz}
\end{eqnarray}
where $\mathscr{H}$ is a generic function of $\left(N,\pi,h_{ij},\pi^{ij};\nabla_{i}\right)$
which, with the second term, corresponds the usual canonical Hamiltonian
and $N^{i}$, $\lambda^{i}$, $\mu_{\text{n}}$, $\nu_{\text{m}}^{i}$
and $\rho_{\text{r}}^{ij}$ play the role of the Lagrange multipliers
corresponding to the following constraints, 
\begin{equation}
\mathcal{H}_{i}\approx0_{i},\quad\pi_{i}\approx0_{i},\label{Int_3d_diff}
\end{equation}
which are associated with the spatial diffeomorphism with $\mathcal{H}_{i}$
the momentum constraints, and 
\begin{equation}
\mathcal{S}^{\text{n}}\approx0^{\text{n}},\quad\mathcal{V}_{i}^{\text{m}}\approx0_{i}^{\text{m}},\quad\mathcal{T}_{ij}^{\text{r}}\approx0_{ij}^{\text{r}},\label{Int_SVT}
\end{equation}
which denote the introduced scalar-, vector- and (symmetric rank-$2$)
tensor-type ACs with $\mathrm{n}$, $\mathrm{m}$ and $\mathrm{r}$
the corresponding indices, respectively. Here, scalar-, vector- and
tensor-types refer to the transformation properties under the spatial
diffeomorphism generated by $\mathcal{H}_{i}$~\footnote{By using the spatial metric $h_{ij}$, its inverse $h^{ij}$ and the
spatial covariant derivative $\nabla_{i}$, one could decompose a
spatial vector into the transverse and longitudinal parts, a spatial
(symmetric rank-$2$) tensor into the transverse-traceless, traceless-longitudinal
and trace parts, as far as the inverse of the Laplace operator is
unique on the spatial manifold. In particular, the decomposition of
a spatial (symmetric rank-$2$) tensor into the traceless and trace
parts does not introduce any non-locality and thus is easy to adopt.
Nonetheless, we shall not employ such decompositions since the main
conclusion of the present paper does not change. Hereafter, we thus
simply classify ACs under the transformation properties under the
spatial diffeomorphism.}. Throughout this work, ``$\approx$'' represents ``weak equality''
that holds only on the constrained subspace $\Gamma_{\text{C}}$ of
the phase space.

The terminology ``primary constraint'' is usually referred to constraints
due to a singular Lagrangian, from which we cannot solve all the conjugate
momenta. In particular, in the case of GR or general SC theories,
in (\ref{Int_3d_diff}) $\pi_{i}\approx0_{i}$ are the primary constraints
due to the absence of the velocity of the shift vector $N^{i}$ in
the Lagrangian, while $\mathcal{H}_{i}\approx0_{i}$ are the so-called
secondary constraints, which arise after making use of the equations
of motion. In this work, since we start from the Hamiltonian in the
phase space from the beginning, a ``primary constraint'' is merely
referred to a constraint that is introduced by hand when defining
the total Hamiltonian. In this sense, both $\pi_{i}\approx0_{i}$
and $\mathcal{H}_{i}\approx0_{i}$, as well as constraints in (\ref{Int_SVT}),
are treated as primary constraints in this work.

We now make some comments on the above introduced constraints. First,
as explained in sec. \ref{Sec_Intro}, in light of the restriction
by the Lovelock's theorem, in order to enlarge the space of theories
such that there is space for searching for the MMG theories, we reduce
the symmetries of the theory from the general covariance to the spatial
covariance. This requires that the spatial diffeomorphism constraints
(\ref{Int_3d_diff}) must be present in the total Hamiltonian (\ref{Int_HT_anz}).
For convenience, we will adopt the extended definition for the momentum
constraints \citep{Gao:2014fra,Mukohyama:2015gia,Saitou:2016lvb,Gao2019}
\begin{eqnarray}
 &  & \mathcal{H}_{i}\equiv-2\sqrt{h}\nabla_{j}\frac{\pi_{i}^{j}}{\sqrt{h}}+\pi\nabla_{i}N\nonumber \\
 &  & +\pi_{j}\nabla_{i}N^{j}+\sqrt{h}\nabla_{j}\frac{\pi_{i}N^{j}}{\sqrt{h}}\approx0_{i},\label{Int_Hi}
\end{eqnarray}
which satisfy the following property \citep{Gao2019} 
\begin{equation}
\left[\mathcal{H}_{i}\left(\vec{x}\right),Q\left(\vec{y}\right)\right]\approx0_{i},\quad\forall Q\approx0,\label{Int_=00003D00003D00005BHi,Q=00003D00003D00005D}
\end{equation}
with $Q$ an arbitrary quantity weakly vanishing on the constrained
phase space $\Gamma_{\text{C}}$. The Poisson bracket $\left[\cdot,\cdot\right]$
is defined by 
\begin{eqnarray}
\left[\mathcal{F},\mathcal{G}\right] & \equiv & \int\text{d}^{3}z\sum_{I}\Big(\frac{\delta\mathcal{F}}{\delta\Phi_{I}\left(\vec{z}\right)}\frac{\delta\mathcal{G}}{\delta\Pi^{I}\left(\vec{z}\right)}\nonumber \\
 &  & -\frac{\delta\mathcal{F}}{\delta\Pi^{I}\left(\vec{z}\right)}\frac{\delta\mathcal{G}}{\delta\Phi_{I}\left(\vec{z}\right)}\Big),\label{Int_PB}
\end{eqnarray}
where we formally denote the ADM variables with $\Phi_{I}$ and their
conjugate momenta with $\Pi^{I}$. By using the property in (\ref{Int_=00003D00003D00005BHi,Q=00003D00003D00005D}),
it is easy to show that the constraints in (\ref{Int_3d_diff}), i.e.~the
spatial diffeomorphism generators, are explicitly classified as the
first-class in terms of Dirac's terminology. They eliminate 12 dimensions
from the (in total) 20-dimensional phase space, leaving us with 8
dimensions at each point of the spacetime, i.e. four local d.o.f.

Second, for the purpose of obtaining the MMG theories, one or several
additional constraints are needed in order to reduce the number of
d.o.f. from four to two. In this work, we perform this by introducing
multiple ACs (\ref{Int_SVT}) to the total Hamiltonian (\ref{Int_HT_anz})
as part of the primary constraints of the theory. Our previous work
\citep{Yao2021}, in which we introduced only one scalar-type AC with
the assumption of a non-dynamical lapse function, can be considered
a preliminary work in that regard. In principle, there could be more
than one AC appearing in the Hamiltonian, including a priori all types
of constraints, i.e.~scalar $\mathcal{S}^{\text{n}}$, vector $\mathcal{V}_{i}^{\text{m}}$
and tensor $\mathcal{T}_{ij}^{\text{r}}$, which are generic functions
of $\left(N,\pi,h_{ij},\pi^{ij};\nabla_{i}\right)$ with the indices
$\text{n}$, $\text{m}$ and $\text{r}$ labeling the number of each
type of AC respectively. Note that it is fair to assume that the ACs
are linearly independent from each other in order to avoid unnecessary
complexity. Obviously, there must be an upper limit of the number
of ACs, otherwise the system will become over-constrained and inconsistent
physically even without coupling to any external fields. The maximum
number of each type of ACs can be obtained in a scenario where all
of the ACs (\ref{Int_SVT}) are assumed to be of the second-class.
In this case, we count the number of d.o.f. as 
\begin{eqnarray}
 &  & \#_{\mathrm{dof}}=\frac{1}{2}\left(\#_{\text{var}}\times2-\#_{\mathrm{1st}}\times2-\#_{\mathrm{2nd}}\right)\nonumber \\
 &  & =\frac{1}{2}\Big[\left(4_{\text{s}}+4_{\text{v}}+2_{\text{t}}\right)\times2-\left(1_{\text{s}}+2_{\text{v}}\right)\times2\times2\nonumber \\
 &  & -1_{\text{s}}\times\mathcal{N}-\left(1_{\text{s}}+2_{\text{v}}\right)\times\mathcal{M}-\left(2_{\text{s}}+2_{\text{v}}+2_{\text{t}}\right)\times\mathcal{R}\Big]\nonumber \\
 &  & =\left(2_{\text{t}}-\mathcal{R}_{\text{t}}\right)-\left(\mathcal{M}_{\text{v}}+\mathcal{R}_{\text{v}}\right)\nonumber \\
 &  & +\frac{1}{2}\left(4_{\text{s}}-\mathcal{N}_{\text{s}}-\mathcal{M}_{\text{s}}-2\times\mathcal{R}_{\text{s}}\right),\label{Int_=00003D00003D000023dof_1}
\end{eqnarray}
where $\mathcal{N},\mathcal{M},\mathcal{R}$ are the numbers of constraints
in each type, and we use the subscripts $\text{s}$, $\text{v}$ and
$\text{t}$ to denote the scalar, vector and tensor d.o.f. respectively.
The classification of d.o.f. into various types are to be understood
in the sense of spatial diffeomorphism with respect to an arbitrary
background. Since the phase space is spanned by $N$, $N^{i}$, $h_{ij}$
and their conjugate momenta, $N$ will contribute one scalar d.o.f.,
i.e. $1_{\text{s}}$ in (\ref{Int_=00003D00003D000023dof_1}), $N^{i}$
will contribute one scalar and two vectorial d.o.f. accounting for
$1_{\text{s}}+2_{\text{v}}$ and $h_{ij}$ will contribute two scalar,
two vectorial and two tensorial d.o.f., accounting for $2_{\text{s}}+2_{\text{v}}+2_{\text{t}}$.
Together with their conjugate momenta, the dimension of the phase
space is therefore $\left(4_{\text{s}}+4_{\text{v}}+2_{\text{t}}\right)\times2$.
The number of d.o.f.~removed by the ACs (\ref{Int_SVT}) are counted
in the similar way in (\ref{Int_=00003D00003D000023dof_1}). Note
that the tensor-type ACs, $\mathcal{T}_{ij}^{\text{r}}\approx0_{ij}^{\text{r}}$,
are symmetric with respect to the subscripts therefore $\mathcal{T}_{ij}^{\text{r}}\approx0_{ij}^{\text{r}}$
account for $-\left(2_{\text{s}}+2_{\text{v}}+2_{\text{t}}\right)\times\mathcal{R}$
in (\ref{Int_=00003D00003D000023dof_1}). The number of d.o.f. in
each type should not be negative in the absence of external fields
otherwise the theory is physically inconsistent. Therefore, from the
last line of (\ref{Int_=00003D00003D000023dof_1}), we require 
\begin{equation}
2-\mathcal{R}\ge0,\quad\mathcal{M}+\mathcal{R}\le0,
\end{equation}
and 
\begin{equation}
4-\mathcal{N}-\mathcal{M}-2\mathcal{R}\ge0,
\end{equation}
which gives 
\begin{equation}
\mathcal{R}=0,\quad\mathcal{M}=0,\quad\mathcal{N}\le4.
\end{equation}
We conclude that none of the tensor- and vector-type ACs are allowed,
and no more than four scalar-type ACs should be introduced. In the
case with four (necessarily second-class) independent scalar-type
ACs, from (\ref{Int_=00003D00003D000023dof_1}) we see that (\ref{Int_HT_anz})
turns out to be an MMG theory automatically, i.e.~without requiring
any further condition. In the next section, we will confirm this result
again, via a more detailed Hamiltonian analysis accounting for all
the possible classifications of the ACs.

Before getting into the next section, we justify the introduction
of (scalar-type) ACs as follows. Even though the ACs are introduced
by hand in this work, we emphasize that they are nothing but part
of the definition of the MMG theories. In other words, it is not possible
to construct an MMG theory without introducing ACs because without
any ACs, the number of d.o.f. in the theory (\ref{Int_HT_anz}) is
four. Therefore additional constraints are necessary to reduce the
number of d.o.f.. One may consider the possibility that extra constraint
may come from the particular choice of the free function $\mathscr{H}\left(N,\pi,h_{ij},\pi^{ij};\nabla_{i}\right)$
in (\ref{Int_HT_anz}), for example by requiring that the lapse $N$
plays the role of a Lagrange multiplier. However, this also yields
the constraint of $\pi\approx0$, which is actually a typical choice
of the scalar-type AC and has been adopted in the previous related
works \citep{Lin:2017oow,Iyonaga2018,Gao:2019twq,Mukohyama2019,Yao2021}
based on the assumption of a non-dynamical lapse. In GR, $\pi\approx0$
is one of the constraints naturally required by the 4-dimensional
spacetime diffeomorphism. In the more general framework with only
spatial covariance, the lapse function could be dynamical in principle
\citep{Domenech:2015tca,Gao2019,Lin2021}, therefore the conjugate
momentum $\pi$ does not correspond to a constraint in general. From
the viewpoint of the formalism presented here, the constraint $\pi\approx0$
is noting but a specific scalar-type AC, which generates the constraint
originally imposed on $\mathscr{H}$. Another example of the AC in
the literature is the gauge fixing term introduced in \citep{Aoki2018,Aoki:2018brq,DeFelice2020b}.
By fixing the gauge condition, which by itself is of the second class,
the first-class constraint becomes a second class constraint. As a
result, the theory is able to couple with matter consistently \citep{Carballo-Rubio2018,Lin2019,Lin2019a}
(see also \citep{Aoki2020,Aoki2020a,Aoki2021}).

Based on the above discussions, we construct a consistent framework
for searching for the MMG theories in the vacuum by introducing the
ACs as follows: 
\begin{equation}
H_{\text{T}}=\int\text{d}^{3}x\left(\mathscr{H}+\mu_{\text{n}}\mathcal{S}^{\text{n}}+N^{i}\mathcal{H}_{i}+\lambda^{i}\pi_{i}\right),\label{Int_HT_2}
\end{equation}
where $\mathscr{H}$ and $\mathcal{S}^{\text{n}}$ with $\mathrm{n}=1,\cdots,\mathcal{N}(\mathcal{N}\leq4)$
are generic functions of $\left(N,\pi,h_{ij},\pi^{ij};\nabla_{i}\right)$.
Before we start to search for the MMG theories based on (\ref{Int_HT_2}),
it is convenient to split the total Hamiltonian (\ref{Int_HT_2})
into two parts 
\begin{equation}
H_{\text{T}}=H_{\text{D}}+H_{\text{P}},\label{Int_HT=00003D00003D00003DHP+HD}
\end{equation}
where $H_{\text{D}}$ denotes the part of the Hamiltonian corresponding
to the spatial diffeomorphism 
\begin{equation}
H_{\text{D}}\equiv\int\text{d}^{3}x\left(N^{i}\mathcal{H}_{i}+\lambda^{i}\pi_{i}\right),\label{Int_HD}
\end{equation}
and the rest in (\ref{Int_HT_2}) is denoted by 
\begin{equation}
H_{\text{P}}\equiv\int\text{d}^{3}x\left(\mathscr{H}+\mu_{\text{n}}\mathcal{S}^{\text{n}}\right),\label{Int_HP}
\end{equation}
which we dub the ``partial'' Hamiltonian \citep{Yao2021}. Clearly,
$H_{\mathrm{D}}$ is fixed in all the SC gravity theories, and thus
$H_{\mathrm{P}}$ plays the central role in the following discussions
, which has nothing to do with $N^{i}$ and $\pi_{i}$. Indeed, we
are allowed to deduct the $\left\{ N^{i},\pi_{i}\right\} $-sector
from the system in the first place since the spatial diffeomorphism
constraints (\ref{Int_3d_diff}) are retained and considered as the
first-class. Therefore, the specificities of the theory such as the
number of d.o.f.~will be completely encoded in $H_{\text{P}}$ with
an $20-6\times2=8$ dimensional phase space. One can however restore
the neglected part $H_{\text{D}}$ without any difficulty in the following
discussions. With the partial Hamiltonian $H_{\text{P}}$ (\ref{Int_HP}),
the number of d.o.f. of the theory is formally counted as 
\begin{eqnarray}
\#_{\mathrm{dof}} & = & \frac{1}{2}\left[\left(2_{\text{t}}+2_{\text{s}}\right)\times2-\#_{\mathrm{1st}}^{\text{s}}\times2-\#_{\mathrm{2nd}}^{\text{s}}\right]\nonumber \\
 & = & 2_{\text{t}}+\frac{1}{2}\left(4_{\text{s}}-\#_{\mathrm{1st}}^{\text{s}}\times2-\#_{\mathrm{2nd}}^{\text{s}}\right),\label{Int_=00003D00003D000023dof_3}
\end{eqnarray}
where $\#_{\mathrm{1st}}^{\text{s}}$ and $\#_{\mathrm{2nd}}^{\text{s}}$
are the numbers of the first- and the second-class constraints, which
include the primary ACs and the possible secondary constraints generated
from the ACs. Clearly, $\#_{\mathrm{1st}}^{\text{s}}$ and $\#_{\mathrm{2nd}}^{\text{s}}$
should satisfy 
\begin{equation}
\mathcal{N}\le\#_{\mathrm{1st}}^{\text{s}}+\#_{\mathrm{2nd}}^{\text{s}}\le4,\label{Int_=00003D00003D0000231+=00003D00003D0000232}
\end{equation}
and 
\begin{equation}
4-\#_{\mathrm{1st}}^{\text{s}}\times2-\#_{\mathrm{2nd}}^{\text{s}}=0,\label{Int_4-2=00003D00003D0000231-=00003D00003D0000232}
\end{equation}
since, for the sake of having an MMG theory, the scalar d.o.f. should
be completely eliminated. Combining (\ref{Int_=00003D00003D0000231+=00003D00003D0000232})
with (\ref{Int_4-2=00003D00003D0000231-=00003D00003D0000232}), we
are able to exhaust all the possible constraint structures for the
MMG theories case by case, which we will do in the next section.

To conclude this section, by performing a Legendre transformation
of the Hamiltonian (\ref{Int_HT_2}), we get the corresponding formal
form of the action as follows 
\begin{equation}
S=\int\text{d}t\text{d}^{3}x\left[N\left(\pi F+2\pi^{ij}K_{ij}\right)-\mathscr{H}-\mu_{\text{n}}\mathcal{S}^{\text{n}}\right],\label{Int_action}
\end{equation}
in which $\pi$ and $\pi^{ij}$ should be understood as the solutions
of the following canonical equations 
\begin{equation}
NF=\frac{\delta H_{\text{P}}}{\delta\pi}\quad\text{ and }\quad2NK_{ij}=\frac{\delta H_{\text{P}}}{\delta\pi^{ij}}.\label{Int_cano_eq}
\end{equation}
Here, $H_{\text{P}}$ is the partial Hamiltonian defined in (\ref{Int_HP})
and we denote 
\begin{equation}
F\equiv\frac{1}{N}\left(\dot{N}-N^{i}\nabla_{i}N\right),\quad K_{ij}\equiv\frac{1}{2N}\left(\dot{h}_{ij}-2\nabla_{(i}N_{j)}\right),
\end{equation}
which play the roles of velocities of the lapse and spatial metric
in the action and the latter is nothing but the extrinsic curvature.
Here and throughout the paper, the overdot ``$\cdot$'' denotes
a time derivative. The Lagrange multipliers $\mu_{\text{n}}$ in (\ref{Int_action})
are determined as follows. According to the classification of the
ACs, the corresponding $\mu_{\text{n}}$ will be fixed by the consistency
conditions of the ACs or kept as some general functions. From (\ref{Int_action})
with (\ref{Int_cano_eq}), we see that the ACs not only appear in
the action directly but also become part of the canonical equations,
hence will influence how the velocities enter the action. Once we
solve for the momenta $\pi$ and $\pi^{ij}$ as functions of $F$
and $K_{ij}$ and substitute them into (\ref{Int_action}), in principle,
we will find the action corresponding to the Hamiltonian (\ref{Int_HT_2}).

\section{Minimally modified gravity with auxiliary constraint(s) \label{Sec_MMG_MAC}}

In this section, we are going to classify the ACs and their secondary
constraints by solving Eqs.~(\ref{Int_4-2=00003D00003D0000231-=00003D00003D0000232})
with (\ref{Int_=00003D00003D0000231+=00003D00003D0000232}), and then
find out the corresponding conditions needed in order to fully satisfy
the classifications so that we are able to exhaust all the possible
constraint structures for the MMG theories. As mentioned in the last
section, one requires no condition in the case with four second-class
(scalar-type) ACs, so we start the discussion from the case with four
ACs, with more details than the previous section.

\subsection{Case with four auxiliary constraints \label{Subsec_S4} }

Let's start with the case with four ACs, i.e. 
\begin{equation}
H_{\text{P}}\equiv\int\text{d}^{3}x\left(\mathscr{H}+\mu_{\text{n}}\mathcal{S}^{\text{n}}\right)\text{ with }\text{n}=1,2,3,4.\label{S4_HP}
\end{equation}
By solving eq. (\ref{Int_=00003D00003D0000231+=00003D00003D0000232})
with (\ref{Int_4-2=00003D00003D0000231-=00003D00003D0000232}), we
find the unique class 
\begin{equation}
\#_{\mathrm{1st}}^{\text{s}}=0,\quad\#_{\mathrm{2nd}}^{\text{s}}=4,\quad\text{\ensuremath{\left(\text{ID key: IV-0-4}\right)}}
\end{equation}
which implies that the four ACs, $\mathcal{S}^{\text{n}}\approx0$,
must be all of the second-class for the sake of having an MMG theory.
(For convenience, we will give each type of classification an identification
key.We label this case IV-0-4.) This is also consistent with the discussions
in the last section. As a consequence, there are two important properties
to be stressed for this case.

First, as an MMG theory, $\mathscr{H}$ and $\mathcal{S}^{\text{n}}$
in (\ref{S4_HP}) can be arbitrarily chosen as the independent functions
of $\left(N,\pi,h_{ij},\pi^{ij};\nabla_{i}\right)$, since no condition
is required to (\ref{S4_HP}). In the cases with less number of ACs,
however, this is not true in general, some conditions on $\mathscr{H}$
and $\mathcal{S}^{\text{n}}$ are needed to obtain the MMG theories.
This complexity is fortunately evaded in this case because we have
introduced a sufficient number of additional constraints.

Second, the MMG theory (\ref{S4_HP}) with (\ref{Int_HD}) can be
coupled with matter consistently. A common problem for the MMG theories
that include the first-class constraint(s) (in addition to the spatial
diffeomorphism constraints (\ref{Int_3d_diff})) is how to couple
with matter consistently. Naive coupling with matter may change the
constraint structure of the theory and thus make the additional first-class
constraint(s) become the second-class, thus reintroduce the scalar
mode(s) suppressed before. Some strategies for dealing with this problem
have been adopted, for example in \citep{Carballo-Rubio2018,Aoki2018,Aoki:2018brq,Lin2019,Lin2019a},
by making the gauge symmetries of the gravity- and matter-sector match
each other. Owing to the absence of additional first-class constraint
in (\ref{S4_HP}), the gauge symmetry of this theory, (\ref{S4_HP})
with (\ref{Int_HD}), is exactly the spatial diffeomorphism. Thus
the coupling problem is automatically solved as long as the matter
field preserves the spatial covariance as well. For instance, the
total Hamiltonian consistently coupled with a scalar field $\phi$
with its conjugate momentum $p$ can be written as 
\begin{equation}
H_{\text{T}}\equiv\int\text{d}^{3}x\left(\hat{\mathscr{H}}+\mu_{\text{n}}\mathcal{S}^{\text{n}}+N^{i}\hat{\mathcal{H}}_{i}+\lambda^{i}\pi_{i}\right)\label{S4_HT}
\end{equation}
where the momentum constraints are extended to 
\begin{equation}
\hat{\mathcal{H}}_{i}\equiv\mathcal{H}_{i}+p\nabla_{i}\phi\approx0_{i},\label{S4_Hi_h}
\end{equation}
with $\mathcal{H}_{i}$ defined in (\ref{Int_Hi}) and the matter
coupled with gravity through the following generic function 
\begin{equation}
\hat{\mathscr{H}}=\hat{\mathscr{H}}\left(N,\pi,h_{ij},\pi^{ij},\phi,p;\nabla_{i}\right).\label{S4_scrH_h}
\end{equation}
Note that, at the classical level, the Lorentz covariance of the scalar
matter could be enhanced by particular choices of the generic function
$\hat{\mathscr{H}}$ although, at the quantum level, it may be violated
via the loops induced effect from the Lorentz-violating gravitons
\citep{Aoki2020}.

To conclude, we show that (\ref{S4_HP}) is a partial Hamiltonian
of an MMG theory without requiring any condition. It can be used to
couple with the matter consistently therefore (\ref{S4_HT}) provides
an extensive yet simple framework for investigating the cosmological
properties of the MMG theories. As an illustrating example, we will
construct a concrete model (by making some choices for the free functions)
in sec. \ref{Sec_CH_FAC}.

\subsection{Case with three auxiliary constraints \label{Subsec_S3} }

In the case of three ACs, 
\begin{equation}
H_{\text{P}}\equiv\int\text{d}^{3}x\left(\mathscr{H}+\mu_{\text{n}}\mathcal{S}^{\text{n}}\right)\text{ with }\text{n}=1,2,3,\label{S3_HP}
\end{equation}
there are two possible classifications according to eq. (\ref{Int_=00003D00003D0000231+=00003D00003D0000232})
with (\ref{Int_4-2=00003D00003D0000231-=00003D00003D0000232}) 
\begin{equation}
\#_{\mathrm{1st}}^{\text{s}}=1,\quad\#_{\mathrm{2nd}}^{\text{s}}=2,\quad\text{\ensuremath{\left(\text{ID key: III-1-2}\right)}}\label{S3_=00003D00003D0000231=00003D00003D00003D1,=00003D00003D0000232=00003D00003D00003D2}
\end{equation}
or 
\begin{equation}
\#_{\mathrm{1st}}^{\text{s}}=0,\quad\#_{\mathrm{2nd}}^{\text{s}}=4.\quad\text{\ensuremath{\left(\text{ID key: III-0-4}\right)}}\label{S3_=00003D00003D0000231=00003D00003D00003D0,=00003D00003D0000232=00003D00003D00003D4}
\end{equation}
The first case (\ref{S3_=00003D00003D0000231=00003D00003D00003D1,=00003D00003D0000232=00003D00003D00003D2}),
i.e.~of type III-1-2, means that one of the ACs in (\ref{S3_HP})
is first-class and the other two are second-class. The existence of
a first-class constraint implies that the ``partial'' Dirac matrix
\begin{equation}
\left[\mathcal{S}^{\text{n}}\left(\vec{x}\right),\mathcal{S}^{\text{n}^{\prime}}\left(\vec{y}\right)\right],\label{S3_DM}
\end{equation}
is degenerate in one dimension and the corresponding linear combination
of $\{\mathcal{S}^{\text{n}}\}$ has vanishing Poisson bracket with
$\mathscr{H}$, which will yield some conditions on the generic functions
$\mathscr{H}$ and $\mathcal{S}^{\text{n}}$. Hence, different from
the case with four ACs, these functions cannot be arbitrarily chosen
anymore.

As a simple case in which the matrix (\ref{S3_DM}) is degenerate
in one dimension, let us suppose 
\begin{equation}
\left[\mathcal{S}^{1}\left(\vec{x}\right),\mathcal{S}^{\text{n}}\left(\vec{y}\right)\right]\approx0^{\text{n}}.\label{S3_MC_1}
\end{equation}
We are always able to make linear combinations among the ACs and redefine
the Lagrange multipliers. So, without loss of generality, we will
continue the discussion with the pattern in (\ref{S3_MC_1}) for its
simplicity. The pattern (\ref{S3_MC_1}) actually defines the conditions
to be imposed on the ACs in order to have an MMG theory. We dub this
kind of conditions the ``minimalizing conditions'' \citep{Yao2021},
since this kind of condition helps to narrow the space of theories
down to the MMG subspace. In other words, the scalar d.o.f.~are completely
eliminated by introducing the ACs satisfying the minimalizing conditions.
Once the minimalizing conditions (\ref{S3_MC_1}) are satisfied, we
should check the consistency condition of $\mathcal{S}^{1}\approx0$,
which is 
\begin{equation}
\dot{\mathcal{S}}^{1}(\vec{x})=\big[\mathcal{S}^{1}(\vec{x}),H_{\text{P}}\big]=\int\text{d}^{3}y\big[\mathcal{S}^{1}(\vec{x}),\mathscr{H}(\vec{y})\big]\approx0.\label{S3_S1_dot}
\end{equation}
According to the classification (\ref{S3_=00003D00003D0000231=00003D00003D00003D1,=00003D00003D0000232=00003D00003D00003D2}),
there should be no secondary constraint generated from (\ref{S3_S1_dot}),
which implies that 
\begin{equation}
\left[\mathcal{S}^{1}\left(\vec{x}\right),\mathscr{H}\left(\vec{y}\right)\right]\approx0,\label{S3_GSC}
\end{equation}
must be satisfied. This is however not a necessary condition for obtaining
an MMG theory. In fact, if eq. (\ref{S3_GSC}) is not satisfied, then
a secondary constraint is generated in (\ref{S3_S1_dot}) and the
system is classified into the class (\ref{S3_=00003D00003D0000231=00003D00003D00003D0,=00003D00003D0000232=00003D00003D00003D4}),
i.e.~of type III-0-4. So no matter whether the condition (\ref{S3_GSC})
is satisfied or not, we will always have an MMG theory as long as
the minimalizing condition (\ref{S3_MC_1}) is satisfied. The condition
(\ref{S3_GSC}) is just used to enhance the gauge symmetry of the
theory without altering the number of d.o.f. We therefore dub this
kind of condition the ``symmetrizing condition''.

We conclude that the partial Hamiltonian (\ref{S3_HP}) describes
an MMG theory as long as the minimalizing condition (\ref{S3_MC_1})
is satisfied and the symmetries of the theory will be enhanced when
the symmetrizing condition (\ref{S3_GSC}) is satisfied. In this case,
the enhanced gauge symmetry helps to suppress the unwanted scalar
d.o.f., which is exactly what happens in GR where the would-be scalar
d.o.f.~is completely suppressed by the spacetime diffeomorphism.
One may ask what kind of gauge symmetry will be enhanced by adopting
the symmetrizing conditions (e.g. (\ref{S3_GSC})) beyond the spatial
diffeomorphism. Some interesting opinions have been discussed in \citep{Lin2019a,Tasinato2020a}.
A detailed analysis is however beyond the scope of the current work.

\subsection{Case with two auxiliary constraints \label{Subsec_S2}}

In the case with two scalar-type ACs, i.e. 
\begin{equation}
H_{\text{P}}\equiv\int\text{d}^{3}x\left(\mathscr{H}+\mu_{\text{n}}\mathcal{S}^{\text{n}}\right)\text{ with }\text{n}=1,2,\label{S2_HP}
\end{equation}
we find the possible classes from eqs. (\ref{Int_=00003D00003D0000231+=00003D00003D0000232})
with (\ref{Int_4-2=00003D00003D0000231-=00003D00003D0000232}) as
follows 
\begin{equation}
\#_{\mathrm{1st}}^{\text{s}}=2,\quad\#_{\mathrm{2nd}}^{\text{s}}=0,\quad\text{\ensuremath{\left(\text{ID key: II-2-0}\right)}}\label{S2_=00003D00003D0000231=00003D00003D00003D2,=00003D00003D0000232=00003D00003D00003D0}
\end{equation}
\begin{equation}
\#_{\mathrm{1st}}^{\text{s}}=1,\quad\#_{\mathrm{2nd}}^{\text{s}}=2,\quad\text{\ensuremath{\left(\text{ID key: II-1-2}\right)}}\label{S2_=00003D00003D0000231=00003D00003D00003D1,=00003D00003D0000232=00003D00003D00003D2}
\end{equation}
and 
\begin{equation}
\#_{\mathrm{1st}}^{\text{s}}=0,\quad\#_{\mathrm{2nd}}^{\text{s}}=4.\quad\text{\ensuremath{\left(\text{ID key: II-0-4}\right)}}\label{S2_=00003D00003D0000231=00003D00003D00003D0,=00003D00003D0000232=00003D00003D00003D2}
\end{equation}
Let's explain the implications of these three classes one by one: 
\begin{enumerate}[label=(\arabic{enumi})]
\item  Obviously, the first class (\ref{S2_=00003D00003D0000231=00003D00003D00003D2,=00003D00003D0000232=00003D00003D00003D0}),
i.e.~type II-2-0, covers the case in which both of the ACs are first-class
without generating secondary constraints. By performing a similar
analysis as the one conducted in the last case (\ref{S3_=00003D00003D0000231=00003D00003D00003D1,=00003D00003D0000232=00003D00003D00003D2}),
it is easy to find the following minimalizing conditions for this
class 
\begin{equation}
\left[\mathcal{S}^{\text{1}}\left(\vec{x}\right),\mathcal{S}^{\text{n}}\left(\vec{y}\right)\right]\approx0^{\text{n}},\label{S2_MC_1-1}
\end{equation}
and 
\begin{equation}
\left[\mathcal{S}^{\text{2}}\left(\vec{x}\right),\mathcal{S}^{\text{2}}\left(\vec{y}\right)\right]\approx0,\label{S2_MC_1-2}
\end{equation}
with the symmetrizing conditions as 
\begin{equation}
\left[\mathcal{S}^{\text{n}}\left(\vec{x}\right),\mathscr{H}\left(\vec{y}\right)\right]\approx0^{\text{n}},\label{S2_GSC_1}
\end{equation}
which prevent the secondary constraints from being generated. 
\item Similarly, the second class (\ref{S2_=00003D00003D0000231=00003D00003D00003D1,=00003D00003D0000232=00003D00003D00003D2}),
i.e.~of type II-1-2, implies that there is one secondary constraint
generated from the two primary constraints and one of these three
constraints is first-class. However, it is easy to show that the secondary
constraint could not be first-class, because then the corresponding
primary constraint would become first-class simultaneously. As a result
there would be two first- and one second-class constraints, which
lead to a negative number of scalar d.o.f.. Therefore, without loss
of generality, we set the secondary constraint as being generated
by the time evolution of $\mathcal{S}^{1}$, i.e., $\dot{\mathcal{S}}^{1}\approx0$.
According to whether the first-class constraint is $\mathcal{S}^{1}\approx0$
or $\mathcal{S}^{2}\approx0$, we divide the class (\ref{S2_=00003D00003D0000231=00003D00003D00003D1,=00003D00003D0000232=00003D00003D00003D2})
into two parallel patterns: 
\begin{enumerate}[label=\alph{enumii})]
\item The first-class constraint is $\mathcal{S}^{1}\approx0$ which yields
the same minimalizing condition as in (\ref{S2_MC_1-1}) but with
\begin{equation}
\left[\mathcal{S}^{1}\left(\vec{x}\right),\dot{\mathcal{S}}^{1}\left(\vec{y}\right)\right]\approx0.\label{S2_MC_2}
\end{equation}
Once (\ref{S2_MC_2}) is satisfied, generally, the consistency condition
of $\dot{\mathcal{S}}^{1}\approx0$ will generate a tertiary constraint
as 
\begin{eqnarray}
 &  & \ddot{\mathcal{S}}^{1}\left(\vec{x}\right)=\int\text{d}^{3}y\Big\{\left[\dot{\mathcal{S}}^{1}\left(\vec{x}\right),\mathscr{H}\left(\vec{y}\right)\right]\nonumber \\
 &  & +\mu_{\text{2}}\left(\vec{y}\right)\left[\dot{\mathcal{S}}^{1}\left(\vec{x}\right),\mathcal{S}^{\text{2}}\left(\vec{y}\right)\right]\Big\}\approx0,\label{S2_ddS1}
\end{eqnarray}
where the Lagrange multiplier $\mu_{2}$ has been fixed by the consistency
condition of $\mathcal{S}^{2}\approx0$. However by requiring (\ref{S2_=00003D00003D0000231=00003D00003D00003D1,=00003D00003D0000232=00003D00003D00003D2}),
$\ddot{\mathcal{S}}^{1}\approx0$ should be prevented from being a
non-trivial constraint therefore (\ref{S2_ddS1}) is forced to be
a symmetrizing condition. We label this case with the identification
key II-1-2a to distinguish it from the next case. 
\item The first-class constraint is $\mathcal{S}^{2}\approx0$, which gives
the same minimalizing conditions (\ref{S2_MC_1-1}) and (\ref{S2_MC_1-2}).
The symmetrizing conditions (\ref{S2_GSC_1}) should be replaced by
\begin{equation}
\left[\mathcal{S}^{2}\left(\vec{x}\right),\dot{\mathcal{S}}^{1}\left(\vec{y}\right)\right]\approx0,\label{S2_GSC_2-1}
\end{equation}
and 
\begin{equation}
\left[\mathcal{S}^{2}\left(\vec{x}\right),\mathscr{H}\left(\vec{y}\right)\right]\approx0,\label{S2_GSC_2-2}
\end{equation}
to ensure that $\mathcal{S}^{2}\approx0$ is first-class and generates
no secondary constraint. Correspondingly, this case is labeled II-1-2b. 
\end{enumerate}
\item The last class (\ref{S2_=00003D00003D0000231=00003D00003D00003D0,=00003D00003D0000232=00003D00003D00003D2}),
i.e.~of type II-0-4, can be simply achieved by giving up the symmetrizing
conditions (\ref{S2_ddS1}), which leads to two primary, one secondary,
and one tertiary constraints, or (\ref{S2_GSC_2-1}) with (\ref{S2_GSC_2-2}),
which leads to two primary and two secondary constraints. All the
resulting constraints are the second-class and both choices require
the same minimalizing conditions (\ref{S2_MC_1-1}), with (\ref{S2_MC_2})
or (\ref{S2_MC_1-2}), respectively. 
\end{enumerate}
We summarize for this case that we find two kinds of minimalizing
conditions, i.e. (\ref{S2_MC_1-1}), with (\ref{S2_MC_1-2}) or (\ref{S2_MC_2})
respectively, for the partial Hamiltonian (\ref{S2_HP}) with two
ACs. For the former, i.e. (\ref{S2_MC_1-1}) with (\ref{S2_MC_1-2}),
one should complement the symmetrizing conditions (\ref{S2_GSC_1}),
or (\ref{S2_GSC_2-1}) with (\ref{S2_GSC_2-2}). In the latter case,
i.e., (\ref{S2_MC_1-1}) with (\ref{S2_MC_2}), one should impose
the symmetrizing condition (\ref{S2_ddS1}). In the previous work,
we have studied a special case of theories in the class (\ref{S2_HP})
in \citep{Yao2021} where $\pi\approx0$ is specifically chosen as
one of the ACs and the particular minimalizing conditions of (\ref{S2_MC_1-1})
and (\ref{S2_MC_2}) for the other AC were also discovered.

\subsection{Case with one auxiliary constraint \label{Subsec_S1}}

As the last case, we study the theory with only one AC, i.e. 
\begin{equation}
H_{\text{P}}\equiv\int\text{d}^{3}x\left(\mathscr{H}+\mu_{\text{1}}\mathcal{S}^{\text{1}}\right),\label{S1_HP}
\end{equation}
which may develop into the same classification as in (\ref{S2_=00003D00003D0000231=00003D00003D00003D2,=00003D00003D0000232=00003D00003D00003D0})-(\ref{S2_=00003D00003D0000231=00003D00003D00003D0,=00003D00003D0000232=00003D00003D00003D2}),
i.e. 
\begin{equation}
\#_{\mathrm{1st}}^{\text{s}}=2,\quad\#_{\mathrm{2nd}}^{\text{s}}=0,\quad\text{\ensuremath{\left(\text{ID key: I-2-0}\right)}}\label{S1_=00003D00003D0000232=00003D00003D0000230}
\end{equation}
\begin{equation}
\#_{\mathrm{1st}}^{\text{s}}=1,\quad\#_{\mathrm{2nd}}^{\text{s}}=2,\quad\text{\ensuremath{\left(\text{ID key: I-1-2}\right)}}\label{S1_=00003D00003D0000231=00003D00003D0000232}
\end{equation}
and 
\begin{equation}
\#_{\mathrm{1st}}^{\text{s}}=0,\quad\#_{\mathrm{2nd}}^{\text{s}}=4.\quad\text{\ensuremath{\left(\text{ID key: I-0-4}\right)}}\label{S1_=00003D00003D0000230=00003D00003D0000234}
\end{equation}
It's obvious in this case that the total number of constraints is
equal to the level of the secondary constraints since $\mathcal{S}^{\text{1}}\approx0$
is the only primary constraint and the secondary constraints must
be generated from it step by step. 
\begin{enumerate}[label=(\arabic{enumi})]
\item  In the first class (\ref{S1_=00003D00003D0000232=00003D00003D0000230}),
i.e.~of type I-2-0, the primary constraint $\mathcal{S}^{\text{1}}\approx0$
and the secondary constraint $\dot{\mathcal{S}}^{1}\approx0$ are
both first-class constraints and it is easy to find the minimalizing
conditions as 
\begin{equation}
\left[\mathcal{S}^{1}\left(\vec{x}\right),\mathcal{S}^{1}\left(\vec{y}\right)\right]\approx0,\label{S1_MC_1-1}
\end{equation}
\begin{equation}
\left[\mathcal{S}^{1}\left(\vec{x}\right),\dot{\mathcal{S}}^{1}\left(\vec{y}\right)\right]\approx0,\label{S1_MC_1-2}
\end{equation}
and 
\begin{equation}
\left[\dot{\mathcal{S}}^{1}\left(\vec{x}\right),\dot{\mathcal{S}}^{1}\left(\vec{y}\right)\right]\approx0,\label{S1_MC_1-3}
\end{equation}
with the symmetrizing condition 
\begin{equation}
\left[\dot{\mathcal{S}}^{1}\left(\vec{x}\right),\mathscr{H}\left(\vec{y}\right)\right]\approx0,\label{S1_GSC_1}
\end{equation}
which prevents further secondary constraints from being generated. 
\item The second class (\ref{S1_=00003D00003D0000231=00003D00003D0000232}),
i.e.~of type I-1-2, implies that only one of the three constraints,
i.e., the primary constraint $\mathcal{S}^{\text{1}}\approx0$, the
secondary constraint $\dot{\mathcal{S}}^{1}\approx0$ and the tertiary
constraint $\ddot{\mathcal{S}}^{1}\approx0$, is first-class. Similarly
to what happens to the class of (\ref{S2_=00003D00003D0000231=00003D00003D00003D1,=00003D00003D0000232=00003D00003D00003D2}),
in this case, the tertiary constraint $\ddot{\mathcal{S}}^{1}\approx0$
is not allowed to be first-class. Therefore, according to whether
the first-class is taken by $\mathcal{S}^{1}\approx0$ or by $\dot{\mathcal{S}}^{1}\approx0$,
we have the following two scenarios for (\ref{S1_=00003D00003D0000231=00003D00003D0000232}): 
\begin{enumerate}[label=\alph{enumii})]
\item  If the first-class constraint is $\mathcal{S}^{1}\approx0$, one
requires the same minimalizing conditions (\ref{S1_MC_1-1}) and (\ref{S1_MC_1-2})
but with 
\begin{equation}
\left[\mathcal{S}^{1}\left(\vec{x}\right),\ddot{\mathcal{S}}^{1}\left(\vec{y}\right)\right]\approx0,\label{S1_MC_2}
\end{equation}
instead of (\ref{S1_MC_1-3}). In order to prevent the generation
of a quaternary constraint we should require the following symmetrizing
condition 
\begin{equation}
\left[\ddot{\mathcal{S}}^{1}\left(\vec{x}\right),\mathscr{H}\left(\vec{y}\right)\right]\approx0.\label{S1_GSC_2}
\end{equation}
We label this case with the identification key I-1-2a to distinguish
it from the next case. 
\item On the other hand, choosing $\dot{\mathcal{S}}^{1}\approx0$ as the
first-class constraint leads exactly to the same minimalizing conditions
as in (\ref{S1_MC_1-1})-(\ref{S1_MC_1-3}) but with a different symmetrizing
condition 
\begin{equation}
\left[\dot{\mathcal{S}}^{1}\left(\vec{x}\right),\ddot{\mathcal{S}}^{1}\left(\vec{y}\right)\right]\approx0.\label{Sn=00003D00003D00003D1_gsc_3}
\end{equation}
This case is labeled with the identification key I-1-2b correspondingly. 
\end{enumerate}
\item The last class (\ref{S1_=00003D00003D0000230=00003D00003D0000234}),
i.e.~of type I-0-4, requires that the primary, secondary, tertiary
and quaternary constraints are all second-class, which yields the
same minimalizing conditions as in case a) i.e., (\ref{S1_MC_1-1}),
(\ref{S1_MC_1-2}) and (\ref{S1_MC_2}), without requiring the symmetrizing
condition. 
\end{enumerate}
In summary, we find two possible kinds of minimalizing conditions
for the partial Hamiltonian (\ref{S1_HP}) with only one AC, both
of which consist in (\ref{S1_MC_1-1}) and (\ref{S1_MC_1-2}), with
(\ref{S1_MC_1-3}) or (\ref{S1_MC_2}) respectively. We can choose
the symmetrizing conditions as either (\ref{S1_GSC_1}) or (\ref{Sn=00003D00003D00003D1_gsc_3})
for the former, and (\ref{S1_GSC_2}) for the latter. We also point
out that these two kinds of minimalizing conditions become equivalent,
i.e. 
\begin{equation}
\left[\dot{\mathcal{S}}^{1}\left(\vec{x}\right),\dot{\mathcal{S}}^{1}\left(\vec{y}\right)\right]=\left[\mathcal{S}^{1}\left(\vec{x}\right),\ddot{\mathcal{S}}^{1}\left(\vec{y}\right)\right]\approx0,\label{S1_=00003D00003D00005BdS,dS=00003D00003D00005D=00003D00003D00003D=00003D00003D00005BS,ddS=00003D00003D00005D}
\end{equation}
when (\ref{S1_MC_1-2}) is satisfied strongly 
\begin{equation}
\left[\mathcal{S}^{1}\left(\vec{x}\right),\dot{\mathcal{S}}^{1}\left(\vec{y}\right)\right]=0.\label{S1_=00003D00003D00005BS1,dS1=00003D00003D00005D}
\end{equation}
This is just what happened in \citep{Mukohyama2019} where $\mathcal{S}^{\text{1}}\approx0$
and $\mathscr{H}$ are respectively specified as $\pi\approx0$ and
\begin{equation}
\mathscr{H}=\mathcal{V}+N\mathcal{H}_{0},
\end{equation}
where $\mathcal{V}$ and $\mathcal{H}_{0}$ are two general functions
of $\left(h_{ij},\pi^{ij};\nabla_{i}\right)$ so that both of the
minimalizing conditions (\ref{S1_MC_1-1}) and (\ref{S1_MC_1-2})
are automatically satisfied strongly and (\ref{S1_=00003D00003D00005BdS,dS=00003D00003D00005D=00003D00003D00003D=00003D00003D00005BS,ddS=00003D00003D00005D})
reduces to 
\begin{equation}
\left[\mathcal{H}_{0}\left(\vec{x}\right),\mathcal{H}_{0}\left(\vec{y}\right)\right]\approx0.
\end{equation}

With all the cases studied above, we have exhausted all of the possible
constraint structures for the MMG theories with AC(s) and found the
corresponding minimalizing and symmetrizing conditions for each class.
For convenience, we summarize the results of subsec.~\ref{Subsec_S4}
to \ref{Subsec_S1} in the table in appendix \ref{App_summary}. In
the next section, as an illustrating example for our formalism, we
will show how to construct a concrete MMG model corresponding to the
case with four ACs by using the generalized Cayley-Hamilton theorem.
We also describe some known MMG theories in appendix \ref{App_examples}
as some concrete examples for the different classifications listed
in tab.~\ref{tab_summary} to better illustrate our formalism.

\section{The Cayley-Hamilton construction with mixed traces constraints \label{Sec_CH_FAC}}

According to the discussions in subsec.~\ref{Subsec_S4}, the total
Hamiltonian with four ACs describes a broad consistent framework to
construct MMG theories, which reads 
\begin{equation}
H_{\text{T}}=\int\text{d}^{3}x\left(\mathscr{H}+\mu_{\text{n}}\mathcal{S}^{\text{n}}+N^{i}\mathcal{H}_{i}+\lambda^{i}\pi_{i}\right),\label{CH_HT}
\end{equation}
where $\mathscr{H}$ and $\mathcal{S}^{\text{n}}$ $\left(\text{n}=1,2,3,4\right)$
are generic functions of $\left(N,\pi,h_{ij},\pi^{ij};\nabla_{i}\right)$,
of which the forms can be taken arbitrarily. This theory is able to
couple with matter consistently and in a general manner, for example
in (\ref{S4_scrH_h}). In order to pick out a concrete MMG model from
this class of theories, we will choose some restrictions for the generic
functions $\mathscr{H}$ and $\mathcal{S}^{\text{n}}$.

First, for simplicity, we assume that the lapse $N$ is non-dynamical,
as being considered in \citep{Lin:2017oow,Iyonaga2018,Gao:2019twq,Mukohyama2019,Yao2021},
which means its conjugate momentum $\pi$ plays the role of an AC.
We take 
\begin{equation}
\mathcal{S}^{4}=\pi\approx0,
\end{equation}
therefore the remaining functions $\mathscr{H}$ and $\mathcal{S}^{\text{I}}$
$\left(\text{I}=1,2,3\right)$ can be rewritten as generic functions
of $\left(N,h_{ij},\pi^{ij};\nabla_{i}\right)$ on the constrained
hypersurface or by redefinition of $\mu_{4}$.

Next we adopt the same restriction as what was imposed in \citep{Yao2021},
i.e.~that the spatial derivative $\nabla_{i}$ appears in the theory
in terms of the Ricci tensor $R_{ij}$ only. Thus we consider the
$\mathscr{H}$ and $\mathcal{S}^{\text{I}}$ to be generic functions
of $\left(N,R_{ij},\pi^{ij}\right)$ only. A motivation for picking
this restriction is that according to the generalized Cayley-Hamilton
theorem \citep{Mertzios1986} (see also the appendix in \citep{Yao2021}),
$\mathscr{H}$ and $\mathcal{S}^{\text{I}}$ can then be equivalently
recast into generic functions of the traces $\left(N,\mathscr{R}^{\text{I}},\varPi^{\text{I}},\mathscr{Q}^{\text{I}}\right)$
constructed from $R_{ij}$ and $\pi^{ij}$, in which we denote the
traces of $R_{ij}$ by 
\begin{equation}
\mathscr{R}^{\text{\text{I}}}\equiv\left\{ R_{i}^{i},R_{j}^{i}R_{i}^{j},R_{j}^{i}R_{k}^{j}R_{i}^{k}\right\} ,\label{CH_RI}
\end{equation}
the traces of $\pi_{ij}$ by 
\begin{equation}
\varPi^{\text{\text{I}}}\equiv\left\{ \pi_{i}^{i},\pi_{j}^{i}\pi_{i}^{j},\pi_{j}^{i}\pi_{k}^{j}\pi_{i}^{k}\right\} ,\label{CH_PiI}
\end{equation}
and the mixed traces by 
\begin{equation}
\mathscr{Q}^{\text{\text{I}}}\equiv\left\{ R_{j}^{i}\pi_{i}^{j},R_{j}^{i}\pi_{k}^{j}\pi_{i}^{k},R_{j}^{i}R_{k}^{j}\pi_{i}^{k}\right\} .\label{CH_QI}
\end{equation}
In the model constructed in \citep{Yao2021}, the mixed traces terms
(\ref{CH_QI}) were dropped for simplicity. Instead, in the current
work, we use them to construct ACs by choosing 
\begin{equation}
\mathcal{S}^{\text{\text{I}}}=\mathscr{Q}^{\text{\text{I}}}-\mathscr{P}^{\text{\text{I}}}\left(N\right)\approx0,\label{CH_QI-PI}
\end{equation}
where $\mathscr{P}^{\text{\text{I}}}$ are generic functions of $N$.
Please keep in mind that we have complete freedom to determine the
form for the ACs in this theory and each choice may define a different
MMG theory. As a result, if we put together the above choices, we
have picked a concrete MMG model from the class of theories (\ref{CH_HT})
as 
\begin{eqnarray}
 &  & H_{\text{T}}^{(\text{C.H.})}=\int\text{d}^{3}x\Big[\mathscr{H}^{(\text{C.H.})}+N^{i}\mathcal{H}_{i}\nonumber \\
 &  & +\lambda^{i}\pi_{i}+\lambda\pi+\mu_{\text{\text{I}}}\left(\mathscr{Q}^{\text{\text{I}}}-\mathscr{P}^{\text{\text{I}}}\right)\Big],\label{CH_HT_CH}
\end{eqnarray}
where the fourth Lagrange multiplier is written as $\mu_{4}\equiv\lambda$
and $\mathscr{H}^{(\text{C.H.})}$ is a free function of $\left(N,\mathscr{R}^{\text{I}},\varPi^{\text{I}}\right)$
on the constrained hypersurface for this MMG model (\ref{CH_HT_CH}).
We will dub this model the Cayley-Hamilton construction with mixed
traces constraints.

\subsection{The dispersion relation}

In order to investigate the properties of the model (\ref{CH_HT_CH})
in a cosmological setting, we will derive the dispersion relation
of gravitational waves as tensor perturbations on a cosmological background
within this model. First, according to (\ref{Int_action}) and (\ref{Int_cano_eq}),
we can easily obtain the corresponding action of the Hamiltonian (\ref{CH_HT_CH})
as follows 
\begin{equation}
S^{(\text{C.H.})}=\int\text{d}t\text{d}^{3}x\Big[2NK_{ij}\pi^{ij}-\mathscr{H}^{(\text{C.H.})}-\mu_{\text{\text{I}}}\Big(\mathscr{Q}^{\text{\text{I}}}-\mathscr{P}^{\text{\text{I}}}\Big)\Big],\label{CH_S}
\end{equation}
where $\pi^{ij}$ should be understood as the solution of 
\begin{equation}
2NK_{ij}=\frac{\partial\mathscr{H}^{(\text{C.H.})}}{\partial\pi^{ij}}+\mu_{\text{I}}\frac{\partial\mathscr{Q}^{\text{I}}}{\partial\pi^{ij}},\label{CH_Can}
\end{equation}
and will rely on the concrete choice of $\mathscr{H}^{(\text{C.H.})}$.
The Lagrange multipliers $\mu_{\text{I}}$ have been fixed by the
consistency condition of $\pi\approx0$ as 
\begin{equation}
\mu_{\text{I}}=\left(\frac{\partial\mathscr{P}^{\text{I}}}{\partial N}\right)^{-1}\frac{\partial\mathscr{H}^{(\text{C.H.})}}{\partial N}.\label{CH_muI}
\end{equation}
The tensor perturbations of the action (\ref{CH_S}) are defined around
a flat Friedmann-Lemaître-Robertson-Walker (FLRW) background with
\begin{equation}
\mu_{\text{I}}=\bar{\mu}_{\text{I}}\left(t\right),\quad N=1,\quad N^{i}=0,\quad h_{ij}=a\left(t\right)^{2}\mathfrak{g}_{ij},\label{CH_BG_Var}
\end{equation}
where (and also throughout the rest of this paper) a bar`` $\bar{}$
'' represents the background values and $a\left(t\right)$ is the
scale factor of the background FLRW metric 
\begin{equation}
\text{d}s^{2}=-\text{d}t^{2}+a\left(t\right)^{2}\delta_{ij}\text{d}x^{i}\text{d}x^{j}.
\end{equation}
We expand 
\begin{equation}
\mathfrak{g}_{ij}\equiv\delta_{ij}+\gamma_{ij}+\frac{1}{2!}\gamma_{ik}\gamma^{k}{}_{j}+\frac{1}{3!}\gamma_{ik}\gamma^{k}{}_{l}\gamma^{l}{}_{j}+\cdots,\label{CH_gij}
\end{equation}
with the tensor perturbation $\gamma^{i}{}_{j}$ satisfying the transverse
and traceless conditions 
\begin{equation}
\partial_{i}\gamma^{i}{}_{j}=0,\quad\gamma_{i}^{i}=0.\label{CH_TT_Gau}
\end{equation}
Note that we have turned off the scalar- and vector-type perturbations
and in this subsection, spatial indices are raised and lowered by
$\delta^{ij}$ and $\delta_{ij}$.

For generality, we keep $\mathscr{H}^{(\text{C.H.})}$ as a general
function. However this also means that we are only able to solve (\ref{CH_Can})
order by order for $\pi^{ij}$. By substituting this solution back
into the action (\ref{CH_S}), we can nonetheless find the following
quadratic action 
\begin{eqnarray}
 &  & S_{2}^{(\text{C.H.})}=\int\text{d}t\text{d}^{3}x\frac{1}{4}\Big(\mathcal{G}_{0}\left(t\right)\dot{\gamma}_{ij}\dot{\gamma}^{ij}\nonumber \\
 &  & +\mathcal{W}_{0}\left(t\right)\gamma_{ij}\frac{\Delta}{a^{2}}\gamma^{ij}-\mathcal{W}_{2}\left(t\right)\gamma_{ij}\frac{\Delta^{2}}{a^{4}}\gamma^{ij}\Big),\label{CH_S_qua}
\end{eqnarray}
where 
\begin{equation}
\mathcal{G}_{0}\left(t\right)\equiv\left[\left(\frac{\partial\mathscr{\bar{H}}}{\partial\varPi^{2}}\right)^{2}-3\frac{\partial\mathscr{\bar{H}}}{\partial\varPi^{3}}\left(\frac{\partial\mathscr{\bar{H}}}{\partial\varPi^{1}}-2H\right)\right]^{-1/2},\label{CH_G0}
\end{equation}
\begin{equation}
\mathcal{W}_{0}\left(t\right)\equiv-\frac{\partial\bar{\mathscr{H}}}{\partial\mathscr{R}^{1}}+\varpi_{0}\left(t\right),
\end{equation}
and 
\begin{equation}
\mathcal{W}_{2}\left(t\right)\equiv\frac{\partial\bar{\mathscr{H}}}{\partial\mathscr{R}^{2}}+\varpi_{2}\left(t\right),
\end{equation}
with 
\begin{eqnarray}
 &  & \varpi_{0}\left(t\right)\equiv-\frac{1}{2}\mathcal{G}_{0}\dot{\bar{\mu}}_{1}+\left(3\frac{\partial\bar{\mathscr{H}}}{\partial\varPi^{3}}\right)^{-1}\left(\mathcal{G}_{0}\frac{\partial\bar{\mathscr{H}}}{\partial\varPi^{2}}-1\right)\dot{\bar{\mu}}_{2}\nonumber \\
 &  & +\left[\left(3\frac{\partial\bar{\mathscr{H}}}{\partial\varPi^{3}}\right)^{-1}\left(\frac{\partial\bar{\mathscr{H}}}{\partial\varPi^{2}}-\mathcal{G}_{0}^{-1}\right)-\frac{\dot{\mathcal{G}}_{0}}{2}+\mathcal{G}_{0}H\right]\bar{\mu}_{1}\nonumber \\
 &  & +\left(3\frac{\partial\bar{\mathscr{H}}}{\partial\varPi^{3}}\mathcal{G}_{0}\right)^{-2}\Big[-1+\mathcal{G}_{0}\Big(3\mathcal{G}_{0}\Big(\mathcal{G}_{0}\frac{\partial\bar{\mathscr{H}}}{\partial\varPi^{3}}\frac{\partial\dot{\bar{\mathscr{H}}}}{\partial\varPi^{2}}\nonumber \\
 &  & +\frac{\partial\dot{\bar{\mathscr{H}}}}{\partial\varPi^{3}}+2\frac{\partial\bar{\mathscr{H}}}{\partial\varPi^{3}}H\Big)+\frac{\partial\bar{\mathscr{H}}}{\partial\varPi^{2}}\Big(2-3\mathcal{G}_{0}\Big(\mathcal{G}_{0}\frac{\partial\dot{\bar{\mathscr{H}}}}{\partial\varPi^{3}}\nonumber \\
 &  & -\frac{\partial\bar{\mathscr{H}}}{\partial\varPi^{3}}\dot{\mathcal{G}}_{0}+2\frac{\partial\bar{\mathscr{H}}}{\partial\varPi^{3}}\mathcal{G}_{0}H\Big)\Big)-\frac{\partial\bar{\mathscr{H}}}{\partial\varPi^{2}}^{2}\mathcal{G}_{0}\Big)\Big]\bar{\mu}_{2},\label{CH_varpi_0}
\end{eqnarray}
and 
\begin{eqnarray}
 &  & \varpi_{2}\left(t\right)\equiv\left(6\frac{\partial\mathscr{\bar{H}}}{\partial\varPi^{3}}\right)^{-2}\Big[12\frac{\partial\mathscr{\bar{H}}}{\partial\varPi^{3}}\left(\mathcal{G}_{0}^{-1}-\frac{\partial\mathscr{\bar{H}}}{\partial\varPi^{2}}\right)\bar{\mu}_{3}\nonumber \\
 &  & -\mathcal{G}_{0}\left(3\frac{\partial\mathscr{\bar{H}}}{\partial\varPi^{3}}\bar{\mu}_{1}+2\left(\mathcal{G}_{0}^{-1}-\frac{\partial\mathscr{\bar{H}}}{\partial\varPi^{2}}\right)\bar{\mu}_{2}\right)^{2}\Big].\label{CH_varpi_2}
\end{eqnarray}
Here, $\mathscr{H}^{(\text{C.H.})}$ is simply denoted by $\mathscr{H}$
for short and $H\equiv\dot{a}/a$ is the Hubble parameter. Note that
the background values of the Lagrange multipliers $\bar{\mu}_{\text{I}}$
in (\ref{CH_varpi_0}) and (\ref{CH_varpi_2}) have been fixed by
(\ref{CH_muI}).

In order to prevent tensor perturbations from being ghosts, i.e.~from
acquiring a negative kinetic term, only the positive branch of $\mathcal{G}_{0}\left(t\right)$
is taken in (\ref{CH_G0}) which holds for $\frac{\partial\mathscr{\bar{H}}}{\partial\varPi^{2}}>0$.
The dispersion relation can be immediately read from (\ref{CH_S_qua})
as \citep{Gao2020} 
\begin{eqnarray}
 &  & \omega_{\text{T}}^{2}=\frac{\mathcal{W}_{0}\left(\tau\right)}{\mathcal{G}_{0}\left(\tau\right)}\frac{k^{2}}{a^{2}}+\frac{\mathcal{W}_{2}\left(\tau\right)}{\mathcal{G}_{0}\left(\tau\right)}\frac{k^{4}}{a^{4}}\nonumber \\
 &  & =\frac{k^{2}}{a^{2}}\mathcal{G}_{0}^{-1}\left[\varpi_{0}-\frac{\partial\bar{\mathscr{H}}}{\partial\mathscr{R}^{1}}+\left(\varpi_{2}+\frac{\partial\bar{\mathscr{H}}}{\partial\mathscr{R}^{2}}\right)\frac{k^{2}}{a^{2}}\right].\label{CH_Dps}
\end{eqnarray}
On large scales, the speed of gravitational waves $c_{\text{T}}=\omega_{\text{T}}/k=1$
when 
\begin{equation}
\frac{\partial\bar{\mathscr{H}}}{\partial\mathscr{R}^{1}}=\varpi_{0}-\mathcal{G}_{0}.\label{CH_cT=00003D00003D00003D1}
\end{equation}
According to the observation of the speed of gravitational waves \citep{Abbott2017,Abbott2017b}
and the modified dispersion relation \citep{Abbott2017a}, we should
impose the following constraints to (\ref{CH_Dps}): 
\begin{equation}
-3\times10^{-15}<\frac{\mathcal{W}_{0}}{\mathcal{G}_{0}}-1<7\times10^{-16},\label{CH_W0/G0}
\end{equation}
and 
\begin{equation}
\left|\frac{\mathcal{W}_{2}}{\mathcal{G}_{0}}\right|<10^{-19}\text{ peV}^{-2},\label{CH_W2/G0}
\end{equation}
where $1\text{ peV}\simeq h\times250\text{ Hz}$ with $h$ the Planck
constant.

\section{Conclusions \label{Sec_cld}}

In this work, we have searched for all the possible Hamiltonian structures
for minimally modified gravity (MMG) theories with multiple auxiliary
constraints (ACs) in the phase space. To do this, we have first investigated
the possibilities of reducing the number of degree(s) of freedom (d.o.f.)
by introducing ACs to the total Hamiltonian (\ref{Int_HT_anz}) while
respecting spatial diffeomorphism. An arbitrary number of scalar-,
vector- and tensor-type ACs have been considered a priori, and in
order to extract the maximum number for each type of ACs, they were
firstly assumed to be linearly independent second-class constraints
following Dirac's terminology. By counting the number of each type
of d.o.f.~with respect to an arbitrary background in (\ref{Int_=00003D00003D000023dof_1}),
and requiring that this number be non-negative, we have found that
no vector- and no tensor-type should be introduced, and that there
should be no more than four scalar-type of ACs in the absence of external
fields~\footnote{This conclusion holds as far as we seek theories with only two tensorial
d.o.f., even if we allow the traceless or transverse-traceless parts
of tensor-type ACs (or the transverse parts of vector-type ACs) to
be imposed separately.}. In fact, the vectorial d.o.f.~have been completely eliminated by
the spatial diffeomorphism constraints (\ref{Int_3d_diff}) and the
existence of vector- or tensor-type ACs would lead to a negative result,
which is physically inconsistent in the vacuum. Hence, on the premise
of retaining spatial covariance, we have determined a consistent framework
(\ref{Int_HT_2}) with no more than four (scalar-type) ACs, $\mathcal{S}^{\text{n}}\approx0^{\text{n}}$
($\mathrm{n}=1,\cdots,\mathcal{N}(\mathcal{N}\leq4)$), as our starting
point for searching for MMG theories which propagate only two tensorial
d.o.f.. By this request, the residual scalar d.o.f.~should all be
completely suppressed by the ACs and the possible secondary constraints
being generated from the consistency conditions of the ACs. According
to the number of introduced primary ACs, we have exhausted all possible
classes of primary and secondary constraints and have found out the
corresponding conditions for their consistency.

In subsec.~\ref{Subsec_S4} with four ACs (\ref{S4_HP}), we have
confirmed that all of the ACs must be classified as second-class,
and that one requires no extra condition to obtain a MMG theory. Therefore,
(\ref{S4_HP}) together with (\ref{Int_HD}) provide an extensive
framework for us to investigate several MMG theories with four ACs
$\mathcal{S}^{\text{n}}\approx0^{\text{n}}$, thus leaving in addition
to the free function $\mathscr{H}$, four other arbitrary choices
of independent functions of $\left(N,\pi,h_{ij},\pi^{ij};\nabla_{i}\right)$,
i.e.~a lot of freedom in terms of model building. More importantly,
this type of structure can be used to couple with matter consistently
in a very general manner in (\ref{S4_scrH_h}). In the sec.~\ref{Sec_CH_FAC},
as an illustrating example of this type of structure, we have constructed
a practical MMG model by starting from (\ref{CH_HT}). For simplicity,
we have restricted ourselves to the special case of a non-dynamical
lapse which yields $\pi\approx0$ as one of the four ACs. We have
further chosen that the spatial derivative $\nabla_{i}$ should only
appear in the theory in the form of the Ricci tensor $R_{ij}$ and,
according to the generalized Cayley-Hamilton theorem, we were therefore
able to recast $\mathscr{H}$ and $\mathcal{S}^{\text{I}}$ into generic
functions of $\left(N,\mathscr{R}^{\text{\text{I}}},\varPi^{\text{I}},\mathscr{Q}^{\text{\text{I}}}\right)$.
By choosing the mixed traces $\mathscr{Q}^{\text{\text{I}}}$ as the
remaining ACs (\ref{CH_QI-PI}), we were able to pick out an interesting
MMG model (\ref{CH_HT_CH}) which we dub the Cayley-Hamilton construction
with mixed traces constraints. In order to investigate the properties
of this model in the cosmological setting, we have studied the tensor
perturbations of the corresponding action (\ref{CH_S}) on an FLRW
background up to the quadratic order and thereby have derived the
modified dispersion relation (\ref{CH_Dps}) for the gravitational
waves (GW) from which we have found that the speed of GW is unity
on large scales when (\ref{CH_cT=00003D00003D00003D1}) is satisfied.
Besides, in order to prevent the tensor perturbations from being ghosts,
the free function $\mathscr{H}^{(\text{C.H.})}\left(N,\mathscr{R}^{\text{I}},\varPi^{\text{I}}\right)$
in the model must satisfy $\frac{\partial\mathscr{\bar{H}}}{\partial\varPi^{2}}>0$
and the constraints (\ref{CH_W0/G0}) and (\ref{CH_W2/G0}) from observations.

In subsec.~\ref{Subsec_S3} to \ref{Subsec_S1}, we have determined
all possible classes of primary and secondary constraints for the
MMG theories with three, two and one AC(s) respectively. Different
from the case with four ACs, in these cases with less AC(s), minimalizing
conditions are required. These are the sufficient conditions to suppress
the scalar d.o.f.~completely, e.g., (\ref{S3_MC_1}) for the case
with three ACs. In particular, with a non-dynamical lapse, the specific
minimalizing conditions, (\ref{S2_MC_1-1}) with (\ref{S2_MC_1-2})
or (\ref{S2_MC_2}), for the case with two ACs had already been discovered
in \citep{Yao2021}; similarly, the minimalizing conditions (\ref{S1_MC_1-1})
and (\ref{S1_MC_1-2}) with (\ref{S1_MC_1-3}) or (\ref{S1_MC_2})
had already been discovered \citep{Mukohyama2019} respectively. On
the other hand, with fewer ACs, there is room for the appearance of
extra gauge symmetries other than the spatial diffeomorphism retained
from the beginning of the construction. In order to allow for the
extra gauge symmetries, so-called symmetrizing conditions have been
imposed, e.g. (\ref{S3_GSC}) for the case with three ACs. The symmetrizing
conditions of course help to suppress the scalar d.o.f., however,
they are neither necessary nor sufficient conditions for obtaining
a MMG theory because we are always able to fix the gauge symmetries
by simply giving up these symmetrizing conditions. Nevertheless they
are important for investigating the gauge symmetries of MMG theories
\citep{Lin2019a,Tasinato2020a} and it would be interesting to clarify
these properties in a future work. We have summarized the minimalizing
and symmetrizing conditions for each class in tab. \ref{tab_summary}.
With this, we have exhausted all the possible constraint structures
of the MMG theories with multiple ACs. To better illustrate our formalism,
we have collected some MMG theories in the appendix \ref{App_examples}
as concrete examples for some of the classifications listed in tab.
\ref{tab_summary}.

We finish this paper with the following comments. First, as mentioned
previously, we should clarify the possible gauge symmetries for the
MMG theories and examine the possible deviations from the spacetime
diffeomorphism. Second, we have not constructed a concrete model for
the case with three ACs which is also an interesting case of MMG theory
and should be studied in the future. Third, the cosmological behavior
and evolution with matter of the Cayley-Hamilton construction with
mixed traces constraints (\ref{CH_HT_CH}) should be investigated
and tested against the observations. It would also be interesting
to see whether it can be used to address current issues within cosmology,
e.g.~the Hubble tension or the dark energy \citep{Ganz2022,Ganz2022a}.
And more importantly, we should find the features of this theory with
respect to GWs and see whether it can be practically distinguished
from GR. Lastly, we comment on the symplectic structure modified by
the ACs. The effects of ACs on the symplectic structure are essentially
the same as what usual constraints do. If all the ACs (and induced
secondary constraints) are classified into the second-class, the induced
two-form has a maximum rank and the second-class ACs can be taken
into account by the use of appropriate Dirac brackets. On the other
hand, in the case with the first-class ACs, the induced two-form is
maximally degenerate and the first-class ACs need to be imposed on
the quantum states. We will come back to these questions in the near
future.

\acknowledgments This work of XG was supported by the National Natural
Science Foundation of China (NSFC) under the grant No. 11975020 and
No. 12005309. The work of SM was supported in part by JSPS Grants-in-Aid
for Scientific Research No.~17H02890, No.~17H06359, and by World
Premier International Research Center Initiative, MEXT, Japan.


\appendix

\section{Summary of the minimalizing and symmetrizing conditions \label{App_summary}}

We summarize the minimalizing and symmetrizing conditions for each
class discussed in sec.~\ref{Subsec_S4} to \ref{Subsec_S1} in tab.~\ref{tab_summary}. 
\begin{widetext}
\begin{table*}
\begin{centering}
\begin{tabular}{|c|c|c|c|c|c|}
\hline 
\# ACs  & Minimalizing condition & Symmetrizing condition  & Classifications  & Identification key  & Examples\tabularnewline
\hline 
$\#^{\text{s}}=4$  & none  & none  & $\#_{\text{1st}}^{\text{s}}=0$, $\#_{\text{2nd}}^{\text{s}}=4$  & IV-0-4  & Mixed Traces\tabularnewline
\hline 
\multirow{2}{*}{$\#^{\text{s}}=3$} & \multirow{2}{*}{$\left[\mathcal{S}^{1},\mathcal{S}^{\text{n}}\right]$} & $\left[\mathcal{S}^{1},\mathscr{H}\right]$  & $\#_{\text{1st}}^{\text{s}}=1$, $\#_{\text{2nd}}^{\text{s}}=2$  & III-1-2  & unknown\tabularnewline
\cline{3-6} \cline{4-6} \cline{5-6} \cline{6-6} 
 &  & none  & $\#_{\text{1st}}^{\text{s}}=0$, $\#_{\text{2nd}}^{\text{s}}=4$  & III-0-4  & unknown\tabularnewline
\hline 
\multirow{6}{*}{$\#^{\text{s}}=2$} & \multirow{4}{*}{$\left[\mathcal{S}^{1},\mathcal{S}^{\text{n}}\right]$ \& $\left[\mathcal{S}^{2},\mathcal{S}^{2}\right]$} & \multirow{2}{*}{$\left[\mathcal{S}^{1},\mathscr{H}\right]$ \& $\left[\mathcal{S}^{\text{2}},\mathscr{H}\right]$} & \multirow{2}{*}{$\#_{\text{1st}}^{\text{s}}=2$, $\#_{\text{2nd}}^{\text{s}}=0$} & \multirow{2}{*}{II-2-0} & \multirow{2}{*}{unknown}\tabularnewline
 &  &  &  &  & \tabularnewline
\cline{3-6} \cline{4-6} \cline{5-6} \cline{6-6} 
 &  & $\left[\mathcal{S}^{2},\dot{\mathcal{S}}^{1}\right]$ \& $\left[\mathcal{S}^{\text{2}},\mathscr{H}\right]$  & $\#_{\text{1st}}^{\text{s}}=1$, $\#_{\text{2nd}}^{\text{s}}=2$  & II-1-2b  & unknown\tabularnewline
\cline{3-6} \cline{4-6} \cline{5-6} \cline{6-6} 
 &  & none  & $\#_{\text{1st}}^{\text{s}}=0$, $\#_{\text{2nd}}^{\text{s}}=4$  & II-0-4b  & Linear AC\tabularnewline
\cline{2-6} \cline{3-6} \cline{4-6} \cline{5-6} \cline{6-6} 
 & \multirow{2}{*}{$\left[\mathcal{S}^{1},\mathcal{S}^{\text{n}}\right]$ \& $\left[\mathcal{S}^{1},\dot{\mathcal{S}}^{1}\right]$} & $\left[\dot{\mathcal{S}}^{1},H_{\text{P}}\right]$  & $\#_{\text{1st}}^{\text{s}}=1$, $\#_{\text{2nd}}^{\text{s}}=2$  & II-1-2a  & 4dEGB\tabularnewline
\cline{3-6} \cline{4-6} \cline{5-6} \cline{6-6} 
 &  & none  & $\#_{\text{1st}}^{\text{s}}=0$, $\#_{\text{2nd}}^{\text{s}}=4$  & II-0-4a  & unknown\tabularnewline
\hline 
\multirow{4}{*}{$\#^{\text{s}}=1$} & \multirow{2}{*}{$\left[\mathcal{S}^{1},\mathcal{S}^{\text{1}}\right]$, $\left[\mathcal{S}^{1},\dot{\mathcal{S}}^{1}\right]$
\& $\left[\dot{\mathcal{S}}^{1},\dot{\mathcal{S}}^{1}\right]$} & $\left[\dot{\mathcal{S}}^{1},\mathscr{H}\right]$  & $\#_{\text{1st}}^{\text{s}}=2$, $\#_{\text{2nd}}^{\text{s}}=0$  & I-2-0  & GR \& $f\left(\mathcal{H}\right)$\tabularnewline
\cline{3-6} \cline{4-6} \cline{5-6} \cline{6-6} 
 &  & $\left[\dot{\mathcal{S}}^{1},\ddot{\mathcal{S}}^{1}\right]$  & $\#_{\text{1st}}^{\text{s}}=1$, $\#_{\text{2nd}}^{\text{s}}=2$  & I-1-2b  & unknown\tabularnewline
\cline{2-6} \cline{3-6} \cline{4-6} \cline{5-6} \cline{6-6} 
 & \multirow{2}{*}{$\left[\mathcal{S}^{1},\mathcal{S}^{\text{1}}\right]$, $\left[\mathcal{S}^{1},\dot{\mathcal{S}}^{1}\right]$
\& $\left[\mathcal{S}^{1},\ddot{\mathcal{S}}^{1}\right]$} & $\left[\ddot{\mathcal{S}}^{1},\mathscr{H}\right]$  & $\#_{\text{1st}}^{\text{s}}=1$, $\#_{\text{2nd}}^{\text{s}}=2$  & I-1-2a  & Cuscuton \& QEC\tabularnewline
\cline{3-6} \cline{4-6} \cline{5-6} \cline{6-6} 
 &  & none  & $\#_{\text{1st}}^{\text{s}}=0$, $\#_{\text{2nd}}^{\text{s}}=4$  & I-0-4  & unknown\tabularnewline
\hline 
\end{tabular}
\par\end{centering}
\begin{centering}
\caption{The minimalizing and symmetrizing conditions.\label{tab_summary}}
\par\end{centering}
\centering{}Note that we simply denote the condition $\left[\cdot\left(\vec{x}\right),\cdot\left(\vec{y}\right)\right]\approx0$
by $\left[\cdot,\cdot\right]$ in the table. 
\end{table*}
\end{widetext}


\section{Some known examples of MMG theories \label{App_examples}}

In this appendix, we collect some known MMG theories as concrete examples
for some of the classifications in tab.~\ref{tab_summary}. Especially,
the lapse functions of the MMG theories collected in this appendix
are all considered to be non-dynamical that, which in our terminology,
implies a specific AC, i.e. $\mathcal{S}^{1}=\pi\approx0$, in the
Hamiltonian of the theories.

\subsection{The QEC gravity}

A model of SCG theory with TTDOF was proposed in the \citep{Gao:2019twq}
whose action is quadratic in the extrinsic curvature (QEC) and by
performing a Legendre transformation we can obtain the total Hamiltonian
of the QEC gravity as 
\begin{equation}
H_{\text{T}}^{\left(\text{QEC}\right)}=\int\text{d}^{3}x\left(\mathcal{V}^{\left(\text{QEC}\right)}+N\mathcal{H}_{0}^{\left(\text{QEC}\right)}+N^{i}\mathcal{H}_{i}+\lambda^{i}\pi_{i}+\lambda\pi\right),\label{QEC_Hcq}
\end{equation}
in which the Hamiltonian constraint of the QEC gravity can be written
as 
\begin{equation}
\mathcal{H}_{0}^{\left(\text{QEC}\right)}\equiv\frac{1}{2\sqrt{h}}\mathcal{G}_{ij,kl}^{\left(\text{W.D.}\right)}\pi^{ij}\pi^{kl}-\sqrt{h}\left[\rho_{1}\left(t\right)+\rho_{2}\left(t\right)R\right],\label{QEC_H0q}
\end{equation}
with the Wheeler-DeWitt metric \citep{DeWitt1967} 
\begin{equation}
\mathcal{G}_{kl,mn}^{\left(\text{W.D.}\right)}\equiv2h_{k(m}h_{n)l}-h_{kl}h_{mn},\label{QEC_G_WD}
\end{equation}
and the part with no lapse is 
\begin{equation}
\mathcal{V}^{\left(\text{QEC}\right)}\equiv\frac{1}{2\sqrt{h}}\mathcal{V}_{ij,kl}^{\left(\text{QEC}\right)}\pi^{ij}\pi^{kl}-\sqrt{h}\left[\rho_{3}\left(t\right)+\rho_{4}\left(t\right)R\right],\label{QEC_Vq}
\end{equation}
with 
\begin{equation}
\mathcal{V}_{kl,mn}^{\left(\text{QEC}\right)}\equiv2\beta_{2}h_{k(m}h_{n)l}-\frac{1}{3}\left(\beta_{1}+2\beta_{2}\right)h_{kl}h_{mn}.\label{QEC_V}
\end{equation}
The coefficients $\beta_{1}$, $\beta_{2}$ and $\rho_{1}$\textasciitilde$\rho_{4}$
are all general functions of time.

From the viewpoint of our formalism, given the total Hamiltonian of
QEC gravity (\ref{QEC_Hcq}), we have 
\begin{equation}
\mathscr{H}=\mathcal{V}^{\left(\text{QEC}\right)}+N\mathcal{H}_{0}^{\left(\text{QEC}\right)},\quad\mathcal{S}^{1}=\pi\approx0,
\end{equation}
and one can check that the minimalizing conditions (\ref{S1_MC_1-1}),
(\ref{S1_MC_1-2}) and (\ref{S1_MC_2}) and the symmetrizing condition
(\ref{S1_GSC_2}) are satisfied, therefore the QEC gravity generally
belongs to the I-1-2a type of MMG theory.

A special case of the QEC gravity is the Cuscuton theory, which corresponds
to 
\begin{equation}
\mathcal{V}^{\left(\text{Cus}\right)}=-\sqrt{h}\mu^{2}\left(t\right).
\end{equation}
In particular, if we set $\mathcal{V}^{\left(\text{QEC}\right)}=0$,
then the symmetrizing condition (\ref{S1_GSC_2}) is trivially satisfied,
which turns the QEC gravity into the I-2-0 type of MMG theory. Especially,
if we further set the coefficient in front of the Ricci scalar to
unity, i.e.~$\rho_{2}=1$, we recover general relativity.

\subsection{The $f\left(\mathcal{H}\right)$ theory}

A more general I-2-0 type MMG theory was proposed in \citep{Carballo-Rubio2018,Mukohyama2019}
whose total Hamiltonian can be written as

\begin{equation}
H_{\text{T}}^{\left(\text{fH}\right)}=\int\text{d}^{3}x\left[Nf\left(\mathcal{H}_{\text{gr}}\right)+N^{i}\mathcal{H}_{i}+\lambda^{i}\pi_{i}+\lambda\pi\right],\label{fH_HT}
\end{equation}
where the Hamiltonian constraint $\mathcal{H}_{0}$ is chosen as a
function of the Hamiltonian constraint in GR, i.e., 
\begin{equation}
\mathcal{H}_{0}^{\left(\text{fH}\right)}\equiv f\left(\mathcal{H}_{\text{gr}}\right),\quad\text{ with }\mathcal{H}_{\text{gr}}\equiv\frac{1}{2\sqrt{h}}\mathcal{G}_{ij,kl}^{\left(\text{W.D.}\right)}\pi^{ij}\pi^{kl}-R,
\end{equation}
the free function $f$ being the reason for it to be dubbed the $f\left(\mathcal{H}\right)$
theory. It's easy to check that the $f\left(\mathcal{H}\right)$ theory
(\ref{fH_HT}) satisfies the minimalizing conditions (\ref{S1_MC_1-1}),
(\ref{S1_MC_1-2}) and (\ref{S1_MC_1-3}) and the symmetrizing condition
(\ref{S1_GSC_1}), which points it to the I-2-0 type of MMG theory
\footnote{As other examples of the I-2-0 type of MMG theories, two different
MMG theories with the square root form of the Hamiltonians were constructed
earlier and dubbed the ``intrinsic time gravity'' in \citep{Murchadha2013}
and ``square root gravity'' in \citep{Lin:2017oow} respectively.}.

\subsection{The consistent $d\rightarrow4$ EGB gravity\label{exam_EGB}}

A concrete example with an additional AC other than $\pi\approx0$
is the consistent $d\rightarrow4$ Einstein-Gauss-Bonnet (4dEGB) gravity
\citep{Aoki2020} in which a gauge fixing condition is introduced
in order to cure the inconsistency of the initial theory \citep{Glavan2020}.
As a result, the total Hamiltonian of the consistent 4dEGB gravity
can be expressed as follows 
\begin{equation}
H_{\text{T}}^{\left(4\text{dEGB}\right)}=\int\text{d}^{3}x\left(N\mathcal{H}_{0}^{\left(4\text{dEGB}\right)}+N^{i}\mathcal{H}_{i}+\lambda^{i}\pi_{i}+\lambda\pi+\mu{}^{3}\mathcal{G}\right),\label{EGB_HT}
\end{equation}
and the Hamiltonian constraint is determined by 
\begin{equation}
\mathcal{H}_{0}^{\left(4\text{dEGB}\right)}\equiv\frac{\sqrt{h}}{2\kappa^{2}}\left[2\Lambda-\mathcal{M}+\tilde{\alpha}\left(4\mathcal{M}_{ij}\mathcal{M}^{ij}-\frac{3}{2}\mathcal{M}^{2}\right)\right],
\end{equation}
where $\kappa$ is the gravitational coupling constant and $\tilde{\alpha}$
is the Gauss-Bonnet coupling with 
\begin{equation}
\mathcal{M}_{ij}\equiv R_{ij}+\mathcal{K}_{k}^{k}\mathcal{K}_{ij}-\mathcal{K}_{ik}\mathcal{K}_{j}^{k}.\label{EGB_M}
\end{equation}
The $\mathcal{K}_{ij}$ in (\ref{EGB_M}) should be understood as
the solution of 
\begin{eqnarray}
 &  & \pi_{j}^{i}=\frac{\sqrt{h}}{2\kappa^{2}}\Big[\mathcal{K}_{j}^{i}-\mathcal{K}\delta_{j}^{i}-\frac{8}{3}\tilde{\alpha}\delta_{jrs}^{ikl}\mathcal{K}_{k}^{r}\nonumber \\
 &  & \times\left(R_{l}^{s}-\frac{1}{4}\delta_{l}^{s}R+\frac{1}{2}\left(\mathcal{M}_{l}^{s}-\frac{1}{4}\delta_{l}^{s}\mathcal{M}\right)\right)\Big],
\end{eqnarray}
with $\delta_{jrs}^{ikl}\equiv3!\delta_{r}^{[i}\delta_{s}^{j}\delta_{t}^{k]}$.
The gauge fixing condition $^{3}\mathcal{G}$ is introduced as a general
function of $\left(h_{ij},\pi^{ij};\nabla_{i}\right)$ and to match
4dEGB gravity (\ref{EGB_HT}) with our framework, we take
\begin{equation}
\mathscr{H}=N\mathcal{H}_{0}^{\left(4\text{dEGB}\right)},\quad\mathcal{S}^{1}=\pi\approx0,\quad\mathcal{S}^{2}={}^{3}\mathcal{G}\approx0.
\end{equation}
One can check that the minimalizing conditions (\ref{S2_MC_1-1})
and (\ref{S2_MC_2}) as well as the symmetrizing condition (\ref{S2_ddS1})
are satisfied, which implies that the consistent 4dEGB gravity (\ref{EGB_HT})
belongs to the II-1-2a type of MMG theory.

A general framework for the II-1-2a type of MMG theory was proposed
in \citep{Yao2021} 
\begin{equation}
H_{\text{T}}=\int\text{d}^{3}x\left(\mathcal{V}+N\mathcal{H}_{0}+N^{i}\mathcal{H}_{i}+\lambda^{i}\pi_{i}+\lambda\pi+\nu\varphi_{0}\right),
\end{equation}
where the free function $\mathcal{V}$, the Hamiltonian constraint
$\mathcal{H}_{0}$ and the AC $\varphi_{0}$ are arbitrary functions
of $\left(h_{ij},\pi^{ij};\nabla_{i}\right)$.

\subsection{The Cayley-Hamilton construction with a linear AC}

In our previous work \citep{Yao2021}, we constructed a concrete MMG
theory with a linear AC by applying the generalized Cayley-Hamilton
theorem as

\begin{equation}
H_{\text{T}}^{\left(\text{LAC}\right)}=\int\text{d}^{3}x\left(\mathscr{H}^{(\text{C.H.})}+N^{i}\mathcal{H}_{i}+\lambda^{i}\pi_{i}+\lambda\pi+\nu\hat{\varphi}\right),\label{LAC_HT}
\end{equation}
where the $\mathscr{H}^{(\text{C.H.})}$ is identical to the free
function in (\ref{CH_HT_CH}) and $\hat{\varphi}$ is called the linear
AC with the following form 
\begin{equation}
\hat{\varphi}\equiv c_{1}\left(t\right)\pi_{i}^{i}+c_{2}\left(t\right)\sqrt{h}R_{i}^{i}+c_{3}\left(t\right)\sqrt{h}\nabla^{2}\frac{\pi_{i}^{i}}{\sqrt{h}}.
\end{equation}
We have demonstrated that the total Hamiltonian (\ref{LAC_HT}) belongs
to the II-0-4b type of MMG theory in \citep{Yao2021} in which we
also proposed a general framework for this type MMG theory as 
\begin{equation}
H_{\text{T}}=\int\text{d}^{3}x\left(\mathscr{H}+N^{i}\mathcal{H}_{i}+\lambda^{i}\pi_{i}+\lambda\pi+\nu\tilde{\varphi}\right),\label{LAC_HTF}
\end{equation}
where the free function $\mathscr{H}$ and the AC $\tilde{\varphi}$
are arbitrary functions of $\left(N,h_{ij},\pi^{ij};\nabla_{i}\right)$
and $\left(h_{ij},\pi^{ij}\right)$ respectively.


\bibliography{MMG_ACF}

\end{document}